\documentclass[twocolumn,prb]{revtex4}
\usepackage{amsmath}
\usepackage{graphicx}
\usepackage{xcolor} 
\usepackage[normalem]{ulem} 

\newcommand{\SB}{\color{black}}
\newcommand{\stkout}[1]{\ifmmode\text{\sout{\ensuremath{#1}}}\else\sout{#1}\fi}
\newcommand{\md}{\mathrm d}

\begin{document}

\title{Skewed Thermodynamic Geometry and Optimal Free Energy Estimation}

\author{Steven Blaber}
\email{sblaber@sfu.ca}
\affiliation{Dept.~of Physics, Simon Fraser University, Burnaby, British Columbia V5A 1S6, Canada}
\author{David A.~Sivak}
\email{dsivak@sfu.ca}
\affiliation{Dept.~of Physics, Simon Fraser University, Burnaby, British Columbia V5A 1S6, Canada}

\date{\today}

\begin{abstract}
Free energy differences are a central quantity of interest in physics, chemistry, and biology. We develop design principles that improve the precision and accuracy of free energy estimators, which has potential applications to screening for targeted drug discovery. {\SB Specifically, by exploiting the connection between the work statistics of time-reversed protocol pairs, we develop near-equilibrium approximations for moments of the excess work and analyze the dominant contributions to the precision and accuracy of standard nonequilibrium free-energy estimators. Within linear response, minimum-dissipation protocols follow geodesics of the Riemannian metric induced by the Stokes' friction tensor. We find the next-order contribution arises from the rank-3 supra-Stokes' tensor that skews the geometric structure such that minimum-dissipation protocols follow geodesics of a generalized cubic Finsler metric. Thus, near equilibrium the supra-Stokes' tensor determines the leading-order contribution to the bias of bidirectional free-energy estimators.}
\end{abstract}


\maketitle

\section{Introduction}
Free energy differences determine the equilibrium phases of thermodynamic systems as well as the relative reaction rates and binding affinities of chemical species.\cite{Gapsys2015} Computational and experimental techniques which accurately and precisely predict free energy differences are therefore highly desirable. One important application is in pharmaceutical drug discovery, where computation of free energy differences can aid in the identification and design of ligands for targeted protein binding.\cite{Schindler2020,Kuhn2017,Ciordia2016,Wang2015} Current methods rely primarily on costly and time-consuming experimentation, which can be reduced through screening with efficient computational techniques.\cite{Schindler2020,Aldeghi2018,Kuhn2017,Ciordia2016,Wang2015,Gapsys2015,Chodera2011,Pohorille2010} 

{\SB Nonequilbrium measurements of free energy differences are often} estimated by measuring the work incurred from a protocol (a dynamic variation in control parameters) that drives the system between control-parameter endpoints corresponding to each state. A unidirectional estimator calculates the free energy difference using only the work done by a protocol driving from initial to final states (forward protocol), while a bidirectional estimator also uses a protocol that drives from final to initial states (reverse protocol). The mean-work estimator~\cite{Gore2003} equates the mean work with the free energy difference and yields a biased estimate for any non-quasistatic (finite-speed) protocol. The Jarzynski estimator for free energy differences (derived from the Jarzynski equality relating the exponentially averaged work to the free energy change~\cite{Jarzynski1997}) is unbiased for a large number of samples. The mean-work and Jarzynski estimators can be used as either uni- or bidirectional estimators; however, if bidirectional data is available the maximum log-likelihood estimate is Bennett's acceptance ratio (BAR)~\cite{Bennett1976} which yields (for a large number of samples) the minimum variance of any unbiased estimator.~\cite{Shirts2003,Maragakis2006}

Regardless of the estimator, non-quasistatic control-parameter protocols result in excess work (work in excess of the free energy difference), which increases the bias and variance (decreases the accuracy and precision, respectively) of free energy estimates. This excess work (equaling the dissipation), and hence the error, can be reduced by following a different path through control-parameter space and varying the velocity along the path while keeping the protocol duration  fixed.~\cite{Reinhardt1992,Hunter1993,Jarque1997} Near equilibrium, this can be mapped onto the thermodynamic-geometry framework: for the mean-work estimator, the bias and variance from finite-time protocols is improved by following geodesics of a thermodynamic metric, the \emph{force-variance} (FV) metric{\SB, a Riemannian metric defined by the covariance matrix of conjugate-force fluctuations};~\cite{Weinhold1975,Salamon1983,Schon1996,Miller2000,Crooks2007} similarly, for BAR, the variance (but not the bias) is minimized if the protocol follows geodesics of the FV metric.~\cite{Shenfeld2009} 
This has been used to improve the precision of calculated binding potentials of mean force.\cite{Minh2019,Pham2011,Pham2012,Park2014}

The FV metric only minimizes the variance of the mean-work estimator and BAR if the relaxation time is independent of the control parameters. The \emph{friction-tensor} metric,~\cite{OptimalPaths,Deffner2020} {\SB the product of the covariance and integral relaxation time of conjugate-force fluctuations~\eqref{force variance}, is more general than the FV and} provides a relatively simple prescription for reducing excess work in more general near-equilibrium processes, where minimum-dissipation protocols follow geodesics of the Riemannian metric induced by the friction tensor. 
For common unidirectional estimators near equilibrium, the bias and variance are proportional to the first moment (mean) of the excess work, and hence minimum-dissipation protocols defined by the friction tensor minimize both the bias and variance. 

We show that for bidirectional estimators, the variance is proportional to the sum of the second moments of the excess work from forward and reverse protocols~\eqref{Variance to second moment}, while the bias is proportional to the difference of the first moments~\eqref{BAR Bias}.\cite{Kim2012} With this in mind, we extend the thermodynamic-geometry framework to higher-order moments of the excess work and beyond the leading-order friction-tensor approximation. We find that for bidirectional estimators near equilibrium, minimum-variance protocols follow geodesics of the Riemannian metric induced by the friction tensor, while minimum-bias protocols follow geodesics of a cubic Finsler metric. For the simple model system of a Brownian particle in a quadratic trap with time-dependent stiffness (a \emph{breathing harmonic trap}), the minimum-variance and minimum-bias protocols can improve variance by a factor of $3-4$ and bias by over a factor of $10$ (Fig.~\ref{Variance_Bias_Ratio}).

\section{Derivation}
Consider a system in the canonical ensemble, in thermal equilibrium with a heat bath at temperature $T$. The probability distribution over microstates $x$ at control parameters $\boldsymbol{\lambda}$ is $\pi(x|\boldsymbol{\lambda}) = \exp{\beta[F(\boldsymbol{\lambda})-E(x,\boldsymbol{\lambda})]}$, with energy $E(x,\boldsymbol{\lambda})$ and free energy $F(\boldsymbol{\lambda}) \equiv -k_{\rm B}T\ln\sum_{x}\exp\left[-\beta E(x,\boldsymbol{\lambda})\right]$. Here $\beta \equiv (k_{\rm B}T)^{-1}$ for Boltzmann's constant $k_{\rm B}$. We define the work done in a single realization of an external agent changing the control parameters $\boldsymbol{\lambda}$ according to a protocol $\Lambda$ as $W \equiv -\int_{0}^{t}\md t' \, f_{i} \dot{\lambda}_{i}(t')$, which implies the average excess work ($W_{\rm ex} \equiv W-\Delta F$) is
\begin{align} 
\langle W_{\rm ex} \rangle_{\Lambda} = -\int_{0}^{t}\md t' \, \langle \delta f_{i} \rangle_{\Lambda}\dot{\lambda}_{i}(t') \ ,
\label{Wex_def}
\end{align}
where we adopt the Einstein summation convention of implied summation over repeated indices. A dot denotes the time derivative $\dot{\lambda}_{i}\equiv \md\lambda_{i}/\md t$, $f_{i} \equiv -\partial_{\lambda_{i}} U$ is the generalized force conjugate to $\lambda_{i}$, and $\delta f_{i} \equiv f_{i} - \langle f_{i} \rangle_{\rm eq}$ is the difference from the equilibrium average. Angle brackets $\langle \cdots\rangle_{\Lambda}$ denote an average over the nonequilibrium ensemble of system responses to control-parameter protocol $\Lambda$. 

The time derivative of the second moment of the excess-work distribution is
\begin{align} 
\frac{\md \langle W_{\rm ex}^{2} \rangle_{\Lambda}}{\md t} = 2\dot{\lambda}_{i}(t) \int_{0}^{t}\md t' \, \left\langle\delta f_{i}(t)\delta f_{j}(t')\right\rangle_{\Lambda}\dot{\lambda}_{j}(t') \ .
\end{align}
For a sufficiently slow protocol, we replace the nonequilibrium average $\langle \cdots\rangle_{\Lambda}$ with the equilibrium average $\langle \cdots\rangle_{\boldsymbol{\lambda}(t)}$ at fixed control parameters $\boldsymbol{\lambda}(t)$:
\begin{align}
\frac{\md \langle W_{\rm ex}^{2} \rangle_{\Lambda}}{\md t} \approx 2\dot{\lambda}_{i}(t)\dot{\lambda}_{j}(t) \int_{0}^{t}\md t'' \, \left\langle\delta f_{i}(0)\delta f_{j}(t'')\right\rangle_{\boldsymbol{\lambda}(t)} \ ,
\end{align}
where we used the stationarity of the equilibrium average, defined $t'' \equiv t' -t$, and assumed smooth protocols to expand the control-parameter velocity to zeroth order, $\dot{\lambda}(t-t'') \approx \dot{\lambda}(t)$. Finally, we assume correlations in the conjugate forces relax quickly relative to the protocol duration and replace the integration bound $t$ with $\infty$ (Appendix~\ref{Finite Bounds}), simplifying the approximation to
\begin{align}
\frac{\md \langle W_{\rm ex}^{2} \rangle_{\Lambda}}{\md t} \approx \frac{2}{\beta}\, \zeta_{ij}^{(1)}[\boldsymbol{\lambda}(t)]\dot{\lambda}_{i}(t)\dot{\lambda}_{j}(t) \ ,
\label{Second moment Friction Approximation}
\end{align}
for the friction tensor
\begin{align}
&\zeta_{ij}^{(1)}[\boldsymbol{\lambda}(t)]  \equiv \beta\int_{0}^{\infty}\md t''\left\langle\delta f_{i}(0)\delta f_{j}(t'')\right\rangle_{\boldsymbol{\lambda}(t)}  \ .
\label{rank 2 friction}
\end{align}
In analogy with fluid dynamics, this rank-two tensor is the \emph{Stokes' friction}, since it produces a drag force that depends linearly on velocity. {\SB We denote the Stokes' friction with superscript (1) since it is the leading-order contribution to dissipation.~\cite{OptimalPaths}}

Following parallel arguments, we approximate the third moment of the excess-work distribution as
\begin{align}
\frac{\md \langle W_{\rm ex}^{3} \rangle_{\Lambda}}{\md t} \approx \frac{3}{\beta^2}\, \zeta_{ijk}^{(2)}[\boldsymbol{\lambda}(t)] \dot{\lambda}_{i}(t)\dot{\lambda}_{j}(t)\dot{\lambda}_{k}(t)  \ ,
\label{Third moment Friction Approximation}
\end{align}
for the rank-three tensor
\begin{align}
\label{rank 3 friction}
&\zeta_{ijk}^{(2)}[\boldsymbol{\lambda}(t)]  \equiv \\
&-\beta^2\int_{0}^{\infty}\md t''\int_{0}^{\infty}\md t'''\left\langle\delta f_{i}(0)\delta f_{j}(t'')\delta f_{k}(t''')\right\rangle_{\boldsymbol{\lambda}(t)}  \ .\nonumber
\end{align}
(The factor of three in \eqref{Third moment Friction Approximation} results from grouping the index permutations $\{ijk,jik,kij\}$ into one term, using the invariance of the sum under exchange of indices, e.g., $\zeta_{ijk}^{(2)}\dot{\lambda}_{i}\dot{\lambda}_{j}\dot{\lambda}_{k} = \zeta_{jik}^{(2)}\dot{\lambda}_{i}\dot{\lambda}_{j}\dot{\lambda}_{k}$.) 
We call the rank-three tensor [Eq.~\eqref{rank 3 friction}] the \emph{supra-Stokes'} tensor {\SB and index it by superscript (2)}, as it corresponds to the leading-order correction to dissipation beyond the Stokes' friction~\eqref{Mean_work}.

For fourth and higher moments, Appendix~\ref{Finite Bounds} shows that the integration bounds must remain finite since the $n$-time covariance functions do not decay to zero, so there is no clear analogy to frictional drag forces. Nevertheless, we can still approximate them by
\begin{align}
\frac{\md \langle W_{\rm ex}^{n} \rangle_{\Lambda}}{\md t} \approx n \mathcal{C}_{\nu_{1}\cdots\nu_{n}}^{(n-1)}[\boldsymbol{\lambda}(t),t] \prod_{i=1}^{n}\dot{\lambda}_{\nu_{i}}(t)  \ ,
\label{Friction Approximation}
\end{align}
where we use index notation $\nu_{1},\nu_{2},\nu_{3}\cdots$ instead of $i,j,k$, and define integral $n$-time covariance functions
\begin{align}
\label{rank n friction}
\mathcal{C}_{\nu_{1}\cdots\nu_{n}}^{(n-1)}&[\boldsymbol{\lambda}(t),t] \equiv \\ 
&(-\beta)^{n}\prod_{i=2}^{n}\int_{0}^{t}\md t_{i}\left\langle\prod_{j=2}^{n}\delta f_{\nu_{1}}(0)\delta f_{\nu_{j}}(t_{j})\right\rangle_{\boldsymbol{\lambda}(t)} \ . \nonumber
\end{align}
This implies that higher-order moments of the excess work are higher order in control-parameter velocity and are therefore smaller for slow protocols. The approximations of \eqref{Second moment Friction Approximation}, \eqref{Third moment Friction Approximation}, and \eqref{Friction Approximation} are the leading-order contributions to each moment of the excess work. 

To derive the next-order contribution we exploit the connection between time-reversed protocols through the Crooks relation,~\cite{Crooks1999,Crooks2000} which constrains the probability of forward and reverse work measurements as
\begin{align}
P_{\Lambda}(W_{\rm ex})e^{-\beta W_{\rm ex}} = {P_{\Lambda^{\dagger}}(-W_{\rm ex})} \ .
\label{Crooks}
\end{align}
$\Lambda^{\dagger}$ is the time-reversed protocol, starting at equilibrium in the end state of the forward protocol. Integrating over $W_{\rm ex}$ produces Jarzynski's equality~\cite{Jarzynski1997}
\begin{align}
\langle e^{-\beta W_{\rm ex}} \rangle_{\Lambda} = 1 \ ,
\label{Jarzynski}
\end{align}
while first multiplying by $W_{\rm ex}$ then integrating leads to
\begin{align}
\langle W_{\rm ex} e^{-\beta W_{\rm ex}} \rangle_{\Lambda} = -\langle W_{\rm ex}  \rangle_{\Lambda^{\dagger}} \ .
\label{eq:Wexp}
\end{align}
Taylor expanding the exponential in \eqref{Jarzynski} and \eqref{eq:Wexp} to third order gives
\begin{subequations}
	\begin{align}
	\langle W_{\rm ex} \rangle_{\Lambda} &\approx \tfrac{1}{2}\beta\langle W_{\rm ex}^2 \rangle_{\Lambda} - \tfrac{1}{6}\beta^2\langle W_{\rm ex}^3 \rangle_{\Lambda} \label{First moment expansion}\\
	\langle W_{\rm ex} \rangle_{\Lambda^{\dagger}} &\approx -\langle W_{\rm ex} \rangle_{\Lambda}+\beta\langle W_{\rm ex}^2 \rangle_{\Lambda} - \tfrac{1}{2}\beta^2\langle W_{\rm ex}^3 \rangle_{\Lambda} \ .
	\end{align}
\end{subequations}
According to \eqref{Friction Approximation}, near equilibrium the higher-order moments of the excess work are higher order in control-parameter velocity and therefore can be neglected for slow protocols. The leading-order contributions to the sum and difference of the excess work are
\begin{subequations}
	\begin{align}
	&\langle W_{\rm ex} \rangle_{\Lambda}+\langle W_{\rm ex} \rangle_{\Lambda^{\dagger}} \approx \beta\langle W_{\rm ex}^2 \rangle_{\Lambda} \label{Sum of first moment}\\
	&\langle W_{\rm ex} \rangle_{\Lambda}-\langle W_{\rm ex} \rangle_{\Lambda^{\dagger}} \approx \tfrac{1}{6}\beta^2\langle W_{\rm ex}^3 \rangle_{\Lambda} \ .
	\label{Difference of first moment}
	\end{align}
\end{subequations}
{\SB Differentiating~\eqref{Sum of first moment} and~\eqref{Difference of first moment} with respect to time, substituting in \eqref{Second moment Friction Approximation} and \eqref{Third moment Friction Approximation} respectively, then adding the two equations gives}
\begin{align}
	\frac{\md \langle W_{\rm ex} \rangle_{\Lambda}}{\md t} &\approx \left(\zeta^{(1)}_{ij}+ \frac{1}{4} \zeta^{(2)}_{ijk}\dot{\lambda}_{k}\right)\dot{\lambda}_{i}\dot{\lambda}_{j} \label{Mean_work} \ ,
\end{align}
where henceforth we drop the explicit dependence of the friction and control-parameter velocity on $\lambda$ and $t$. {\SB Differentiating \eqref{First moment expansion} and substituting \eqref{Mean_work} and \eqref{Third moment Friction Approximation} gives}
\begin{align}
	\frac{\md \langle W_{\rm ex}^2 \rangle_{\Lambda}}{\md t} &\approx \frac{2}{\beta}\left(\zeta^{(1)}_{ij} + \frac{3}{4}\zeta^{(2)}_{ijk}\dot{\lambda}_{k}\right)\dot{\lambda}_{i}\dot{\lambda}_{j} \ . \label{Variance friction approximation}
\end{align}

To derive an approximation for higher-order moments, we follow parallel arguments. Multiplying \eqref{Crooks} by $W_{\rm ex}^n$ and integrating over $W_{\rm ex}$, then either adding or subtracting $\langle (-W_{\rm ex})^{n}\rangle_{\Lambda^{\dagger}}$, the leading-order contributions to the sum and difference for $n>2$ are
\begin{subequations}
	\begin{align}
	&\langle W_{\rm ex}^{n} \rangle_{\Lambda} +\langle (-W_{\rm ex})^{n} \rangle_{\Lambda^{\dagger}} \approx 2\langle (-W_{\rm ex})^{n} \rangle_{\Lambda} \label{Sum of higher moments}\\
	&\langle W_{\rm ex}^{n} \rangle_{\Lambda} - \langle (-W_{\rm ex})^{n} \rangle_{\Lambda^{\dagger}} \approx \beta\langle W_{\rm ex}^{n+1} \rangle_{\Lambda} \ ,
	\label{Difference of higher moments}
	\end{align}
\end{subequations}
which implies
\begin{align}
\beta^{n-1}&\frac{\md \langle W_{\rm ex}^{n} \rangle_{\Lambda}}{\md t} \approx \label{Higher order moment friction approximation} \\
&\left(n\mathcal{C}_{\nu_{1}\cdots\nu_{n}}^{(n-1)} + \frac{n+1}{2}\mathcal{C}_{\nu_{1}\cdots\nu_{n+1}}^{(n)}\dot{\lambda}_{\nu_{n+1}}\right)\prod_{i=1}^{n}\dot{\lambda}_{\nu_{i}} \ . \nonumber
\end{align}
Equations~\eqref{Mean_work}, \eqref{Variance friction approximation}, and \eqref{Higher order moment friction approximation} are the central results of this article, which have applications to designing minimum-dissipation and optimal free energy estimation protocols.

\section{Next-order contribution to excess work}
The approximation of \eqref{Mean_work} can improve near-equilibrium estimates of the excess work. The leading-order term is the usual linear-response approximation (Eq.~(13) of Ref.~\onlinecite{OptimalPaths}) and is positive for all protocols. The Stokes' friction~\eqref{rank 2 friction} is an autocovariance function, 
\begin{align}
\zeta^{(1)}_{ij}  =  \langle\delta f_{i}\delta f_{j}\rangle_{\boldsymbol{\lambda}(t)} \circ \tau_{ij}^{(1)} \ ,
\label{force variance}
\end{align}
the Hadamard (entry-by-entry) product, $\circ$, of the conjugate-force covariance $\langle\delta f_{i}\delta f_{j}\rangle_{\boldsymbol{\lambda}(t)}$ and integral relaxation time
\begin{align} 
\tau_{ij}^{(1)}  \equiv \int_{0}^{\infty}\md t''\frac{\langle\delta f_{i}(0)\delta f_{j}(t'')\rangle_{\boldsymbol{\lambda}(t)}}{\langle \delta f_{i} \delta f_{j}\rangle_{\boldsymbol{\lambda}(t)}} \ .
\end{align}
The Stokes' friction is largest when the conjugate-force fluctuations are largest (large covariance) and most persistent (long relaxation time). In contrast to the positive contribution from the Stokes' friction, the contribution from the supra-Stokes' tensor (second term in \eqref{Mean_work}) changes sign under time reversal because it is cubic in control-parameter velocity. The supra-Stokes' tensor~\eqref{rank 3 friction}
\begin{align}
\label{coskewness}
\zeta_{ijk}^{(2)} = -\langle\delta f_{i}\delta f_{j}\delta f_{k}\rangle_{\boldsymbol{\lambda}(t)}  \circ\tau_{ijk}^{(2)}
\end{align}
is not positive semidefinite, and is the Hadamard product of the unnormalized coskewness $\langle\delta f_{i}\delta f_{j}\delta f_{k}\rangle_{\boldsymbol{\lambda}(t)}$ (related to skewness as covariance is related to variance) and the integral double relaxation time 
\begin{align}
\tau_{ijk}^{(2)}& \equiv \\
&\int_{0}^{\infty}\md t''\int_{0}^{\infty}\md t'''\frac{\langle\delta f_{i}(0)\delta f_{j}(t'')\delta f_{k}(t''')\rangle_{\boldsymbol{\lambda}(t)}}{\langle \delta f_{i} \delta f_{j}\delta f_{k}\rangle_{\boldsymbol{\lambda}(t)}} \ . \nonumber
\end{align}
For a protocol with positive velocity, the contribution to dissipation from the supra-Stokes' tensor quantifies the increase (decrease) in excess work from negatively (positively) skewed conjugate-force fluctuation.

The leading-order friction-tensor approximation (first term in \eqref{Mean_work}) endows the control-parameter space with a Riemannian metric, such that minimum-work protocols follow geodesics of the Stokes' friction tensor. With the addition of the next-order contribution~\eqref{rank 3 friction}, the excess work can be expressed as
\begin{align}
\langle W_{\rm ex} \rangle_{\Lambda} \approx \int_{0}^{\Delta t}\md t \, \zeta_{ij}^{\rm tot}\dot{\lambda}_{i}\dot{\lambda}_{j} \ ,
\label{total excess work}
\end{align}
for total friction tensor
\begin{align}
\zeta_{ij}^{\rm tot} \equiv \zeta^{(1)}_{ij}+\tfrac{1}{4}\zeta^{(2)}_{ijk}\dot{\lambda}_{k}
\label{total friction}
\end{align}
that explicitly depends on the protocol velocity. Minimum-dissipation protocols follow geodesics of the generalized cubic Finsler metric $\zeta_{ij}^{\rm tot}$, an extension of the Riemannian metric $\zeta^{(1)}_{ij}$. {\SB A Finsler metric defines a space where distances depend not only on positions (points) but also on directions (tangent vectors). Nevertheless, the usual concepts of length, curvature, and geodesics from Riemannian geometry generalize, and there are standard procedures for calculating geodesics~\cite{Antonelli2013} (see Appendix~\ref{Finsler Geometry}).} For small control-parameter velocities, the total friction is positive semi-definite; however, for large control-parameter velocities the approximation breaks down and no longer guarantees positive semidefiniteness, leading to the unphysical possibility of negative excess work.

\begin{figure}
	\includegraphics[width=0.5\textwidth]{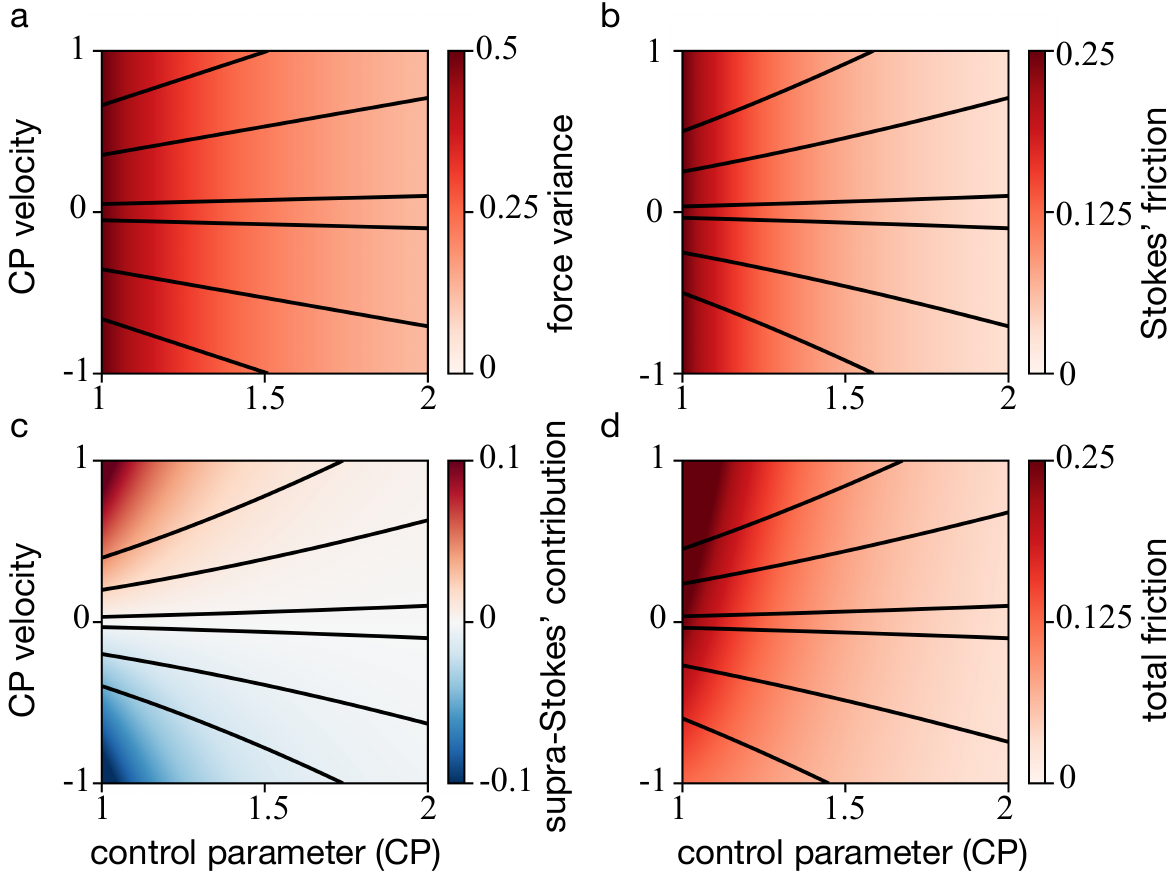}
	\caption{{\bf Friction as a function of control-parameter velocity $\dot{\lambda}^{*}$ and control parameter $\lambda^{*}$	for the breathing harmonic trap.} Scaled (a) force variance $\langle \delta f^{2}\rangle^{*} \equiv \beta^2k_{\rm i}^2\langle \delta f^{2}\rangle$, (b) Stokes' friction $\zeta^{(1)*}\equiv \beta^2k_{\rm i}^3\zeta^{(1)}/\gamma$, (c) supra-Stokes' contribution $\frac{1}{4}\dot{\lambda}^{*}\zeta^{(2)*} \equiv \frac{1}{4}\beta^2k_{\rm i}^3\dot{\lambda}\zeta^{(2)}/\gamma$, and (d) total friction $\zeta^{\rm tot*} \equiv \beta^2k_{\rm i}^3\zeta^{\rm tot}/\gamma$, for initial spring constant $k_{\rm i}$ and damping coefficient $\gamma$. The control parameter is the trap stiffness, given in dimensionless form as $\lambda^{*} \equiv k/k_{\rm i}$ with velocity $\dot{\lambda}^{*} \equiv \gamma k/k_{\rm i}^{2}$. Geodesics (solid curves) minimize the magnitude of the corresponding metric's contribution to the excess work, with distinct curves representing different average protocol velocities.}
	\label{Friction_Metric}
\end{figure}

To illustrate some general properties of the friction, Fig.~\ref{Friction_Metric} shows, for the model system of a \emph{breathing harmonic trap} (Brownian particle in a harmonic trap where the control parameter is the time-dependent stiffness, see Appendix~\ref{Breathing trap}), the (a) force variance, (b) Stokes' friction, (c) supra-Stokes' contribution, and (d) total friction. The force variance and Stokes' friction are independent of control-parameter velocity. In general, the force variance differs from the Stokes' friction whenever integral relaxation time depends on the control parameters, as for the breathing harmonic trap. The contribution from the supra-Stokes' tensor (c) is antisymmetric in control-parameter velocity, becoming negative for negative velocity. This antisymmetric contribution skews the total friction (d), which depends on the control-parameter velocity and lacks any symmetry under time reversal.

Geodesics (solid curves in Fig.~\ref{Friction_Metric}) are protocols that minimize the contribution from the corresponding metric to the excess work. 
For relatively fast protocols (average scaled velocities of $\dot{\lambda}^{*} \equiv k/k_{\rm i} \gtrsim 0.5$ or $\lesssim -0.5$, for initial spring constant $k_{\rm i}$ and damping coefficient $\gamma$), the velocity is significantly smaller in regions of high friction and larger in regions of low friction (or force variance). 

\section{Precision and accuracy of free energy estimates}
We quantify the precision of a free energy estimator by its variance. In the limit of many samples, the expected variance of any unbiased estimator $\widehat{\Delta F}$ is bounded by~\cite{Shirts2003}
\begin{align}
\left\langle \left(\delta \widehat{\Delta F}\right)^2 \right\rangle
\ge \frac{2}{N}\langle[1+\cosh(\beta W_{\rm ex})]^{-1}\rangle^{-1}-\frac{4}{N} \ ,
\label{Variance}
\end{align}
where for simplicity we assume an equal number of forward and reverse work measurements, and the average $\langle \cdots \rangle$ is over a total of $N$ work measurements. BAR saturates this bound for large $N$. This bound demonstrates an explicit connection between the minimum variance of free energy estimators and the excess work. The variance is minimized at $0$ when the excess-work distribution is a delta function at $W_{\rm ex} = 0$ (achieved in the quasistatic limit). For any finite-time protocol, the average excess work is positive.

Assuming small excess work, we expand the variance
\begin{subequations}
	\begin{align}
	\left\langle \left(\delta \widehat{\Delta F}\right)^2 \right\rangle & \approx \frac{\langle W_{\rm ex}^{2}\rangle_{\Lambda} +\langle W_{\rm ex}^{2}\rangle_{\Lambda^{\dagger}}}{2N} \label{Variance to second moment}\\
	&\approx \frac{2}{\beta N}\int_{0}^{\Delta t}\md t \,  \zeta^{(1)}_{ij} \dot{\lambda}_{i}\dot{\lambda}_{j} \ ,
	\label{BAR Variance Friction}
	\end{align}
\end{subequations}
where the second line follows from \eqref{Variance friction approximation}. Equation~\eqref{Variance to second moment} also holds for very few samples, since then BAR is equivalent to the average of the sum of the forward and reverse work measurements.~\cite{Kim2012} Thus the protocol designed to reduce the variance follows geodesics of $\zeta^{(1)}_{ij}$, and for one-dimensional control proceeds at velocity $\dot{\lambda} \propto \left(\zeta^{(1)}\right)^{-1/2}$.

Unlike the variance, the protocol that maximizes the accuracy (minimum-bias) is different for unidirectional and bidirectional estimators. For unidirectional Jarzynski and mean-work estimators, near equilibrium the minimum-bias protocol is simply the minimum-dissipation protocol (protocol that minimizes \eqref{Mean_work}) and therefore to leading order is optimized by the same protocol that minimizes \eqref{BAR Variance Friction}. For BAR, for small excess work (or few samples), the bias instead is proportional to the difference between the forward and reverse excess work:~\cite{Kim2012}
\begin{subequations}
	\begin{align}
	\left\langle \delta \widehat{\Delta F}\right\rangle &\approx \frac{1}{2N}\left(\langle W_{\rm ex}\rangle_{\Lambda}-\langle W_{\rm ex}\rangle_{\Lambda^{\dagger}}\right)\label{BAR Bias} \\
	&\approx \frac{1}{4N}\int_{0}^{\Delta t}\md t \, \zeta^{(2)}_{ijk} \,\dot{\lambda}_{i}\dot{\lambda}_{j}\dot{\lambda}_{k} \ ,
	\label{BAR Bias Friction}
	\end{align}
\end{subequations}
where the second line follows from \eqref{Mean_work}. The protocol designed to reduce the (magnitude of) bias thus follows geodesics of the cubic Finsler metric $\zeta^{(2)}_{ijk}$, simplifying for one-dimensional control to $\dot{\lambda} \propto \left(\zeta^{(2)}\right)^{-1/3}$.

\begin{figure}
	\includegraphics[width=0.5\textwidth]{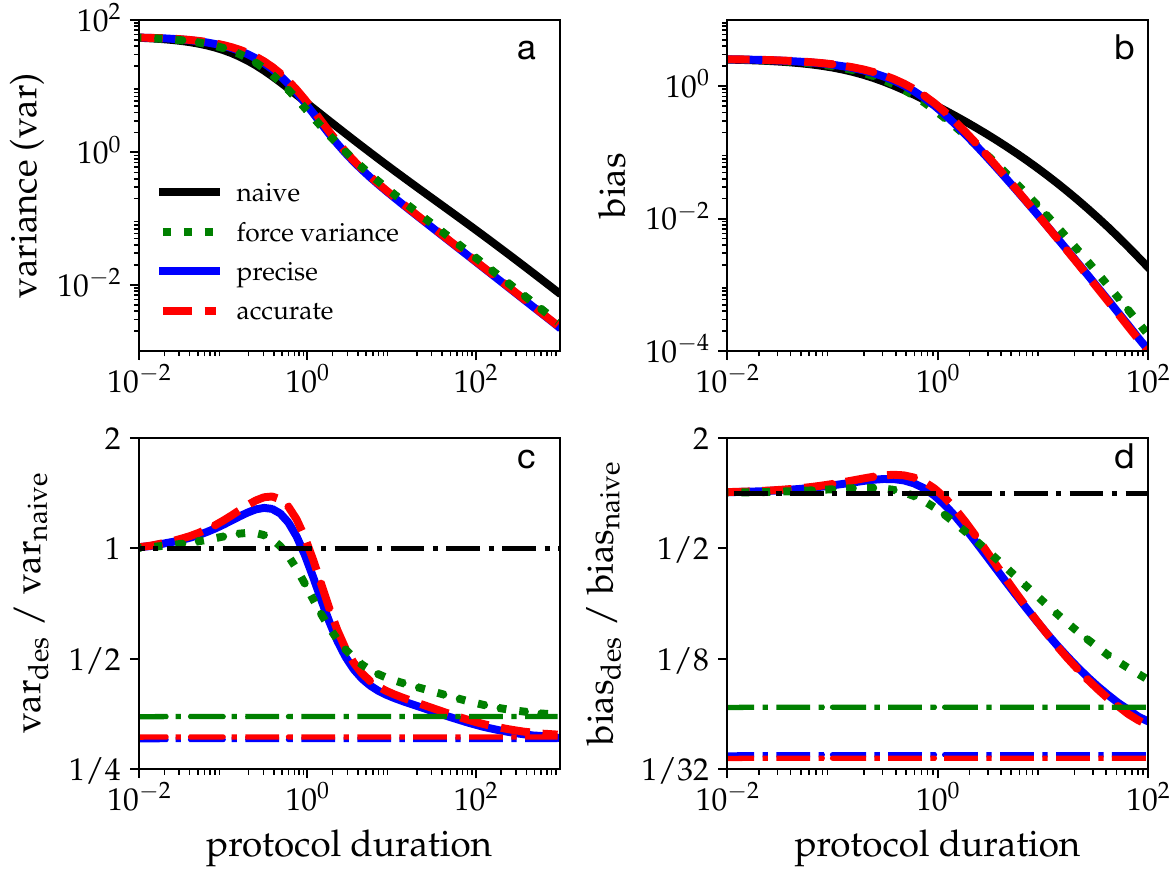}
	\caption{(a) Scaled variance $\langle (\beta\, \delta \widehat{\Delta F})^2 \rangle/N$~\eqref{Variance to second moment} and (b) scaled bias $\langle \beta\, \delta \widehat{\Delta F}\rangle/N$~\eqref{BAR Bias} as functions of scaled protocol duration $\Delta t/\tau^{(1)}_{\rm f}$ (for slowest relaxation time $\tau^{(1)}_{\rm f} = \gamma/(2k_{\rm f})$), for the breathing harmonic trap. Ratios of designed and naive (c) variances $\langle (\delta \widehat{\Delta F})^2\rangle_{\rm des}/\langle (\delta \widehat{\Delta F})^2\rangle_{\rm naive}$ and (d) biases $\langle \delta \widehat{\Delta F}\rangle_{\rm des}/\langle \delta \widehat{\Delta F}\rangle_{\rm naive}$. Subscript `naive' denotes the constant-velocity protocol, and `des' the designed protocols, which are the force-variance-optimized `force variance', the minimum-variance `accurate', and the minimum-bias `precise' protocols. Dash-dotted lines denote the limits for short (black) and long duration (colors). Final control-parameter value (spring constant) is $k_{\rm f}/k_{\rm i}=1/16$.}
	\label{Variance_Bias_Ratio}
\end{figure}

To illustrate the potential benefit of protocols designed to reduce variance and bias,  Fig.~\ref{Variance_Bias_Ratio} shows approximations for the variance~\eqref{Variance to second moment} and bias~\eqref{BAR Bias} of designed and naive protocols, as well as their ratio, as a function of protocol duration, for the model system of a breathing harmonic trap (Appendix~\ref{Breathing trap}). For a slow protocol (protocol duration longer than the slowest relaxation time, $\Delta t/\tau^{(1)}_{\rm f} \gtrsim 1$), the designed protocols reduce the variance by a factor of $3-4$ (Fig.~\ref{Variance_Bias_Ratio}c), with the precise protocol (designed to reduce variance) performing the best (smallest ratio). For the bias, the accurate protocol (designed to reduce bias) performs the best (smallest ratio), with reductions by an order of magnitude for the slowest protocols shown (Fig.~\ref{Variance_Bias_Ratio}d). For fast protocols (protocol duration shorter than the slowest relaxation time, $\Delta t/\tau^{(1)}_{\rm f} \lesssim 1$), the approximations break down, and the naive protocol can outperform the designed in both bias and variance (ratio larger than one). For this system the minimum-variance and minimum-bias protocols achieve similar amounts of bias and variance in all cases, likely due to their similar functional form (Fig.~\ref{Protocol}). 

\section{Discussion\label{Discussion}}
We have developed near-equilibrium approximations for moments of the excess work, \eqref{Mean_work}, \eqref{Variance friction approximation}, and \eqref{Higher order moment friction approximation}, that incorporate time-reversal symmetric and antisymmetric contributions. The antisymmetric contribution to the first moment~\eqref{Mean_work} yields the next-order contribution beyond linear response, which asymmetrically skews the Riemannian metric \eqref{rank 2 friction} into a generalized cubic Finsler metric~\eqref{total friction}. The Stokes' friction tensor~\eqref{rank 2 friction} controls the leading-order contribution to the variance of both unidirectional and bidirectional free energy estimators~\eqref{Variance friction approximation}, such that minimum-variance protocols follow geodesics of the Stokes' friction. For unidirectional estimators, these same protocols also minimize the bias; however, for bidirectional estimators such as BAR the leading-order contribution to the bias is from the supra-Stokes' tensor~\eqref{BAR Bias Friction}, and therefore minimum-bias protocols instead follow geodesics of the supra-Stokes' tensor. From these near-equilibrium approximations, we design protocols that increase the precision and accuracy of standard nonequilibrium free energy estimators (Fig.~\ref{Variance_Bias_Ratio}).

The addition of the supra-Stokes' tensor has several physical implications. Notably, it accounts for time-reversal asymmetric dissipation, which arises from skewed conjugate-force fluctuations. Since the supra-Stokes' contribution is antisymmetric, it cancels out for equal numbers of forward and reverse protocols, which has implications for the design principles of molecular machines: many molecular machines (e.g., kinesin walking toward a microtubule's plus end~\cite{Hirokawa2009} or ATP synthase synthesizing ATP~\cite{Junge2015}) achieve directed behavior; the supra-Stokes' tensor quantifies the leading-order energetic cost of directed operation compared to a coequal forward and reverse process.

In some cases the minimum-work protocols can be calculated exactly,~\cite{Schmiedl2007,Gomez2008} or minimum-variance protocols solved numerically,~\cite{Geiger2010,Solon2018} which would yield more accurate and precise estimators; however, to date these optimizations are limited to simple systems and do not permit straightforward generalization. The present formalism can be applied to more general settings than exact calculations and makes fewer approximations than Shenfeld \emph{et al.},~\cite{Shenfeld2009} who only examine the force-variance metric and do not consider minimizing bias.

Here we considered continuous protocols, but free energies are often estimated from sampling discrete states.\cite{Pearlman1989,Lu1999,Radmer1997,Wu2005} Additionally, for discrete state sampling in replica-exchange~\cite{Sugita2000} or parallel-tempering~\cite{Earl2005} simulations it is important to consider the acceptance probability between states.\cite{Kofke2002,Predescu2004,Kone2005,Rathore2005,Predescu2005,Nadler2007,Chodera2011Replica,Dirks2012,Maccallum2018} The force-variance metric has been applied in this context,~\cite{Shenfeld2009} and the linear-response framework of Ref.~\onlinecite{OptimalPaths} has been generalized to discrete control.~\cite{Large2019} As Ref.~\onlinecite{Shenfeld2009} has shown, spacing the states along geodesics of the force variance not only reduces the variance of the free energy estimator but also increases the acceptance probability. We expect that analogous principles hold for designing minimum-variance and minimum-bias protocols for discrete state sampling using the milder approximations of our approach.

\begin{acknowledgments}
	The authors thank Steve Large, Emma Lathouwers, and Joseph Lucero (SFU Physics), Miranda Louwerse (SFU Chemistry), John Chodera and Josh Fass (Memorial Sloan Kettering Computational Biology), and Gavin Crooks (Berkeley Institute for Theoretical Science) for feedback on the manuscript. This work is supported by an SFU Graduate Deans Entrance Scholarship (SB), an NSERC Discovery Grant and Discovery Accelerator Supplement (DAS), and a Tier-II Canada Research Chair (DAS). 
\end{acknowledgments}

\section*{Data Availability}
The data that support the findings of this study are available from the corresponding author upon reasonable request.

\appendix
\section{Breathing Harmonic Trap}
\label{Breathing trap}

A simple non-trivial system for designing and testing minimum-variance and minimum-bias protocols is the breathing harmonic trap, since its correlation functions can be calculated analytically and it has a non-Gaussian work distribution. Consider a colloidal particle in a harmonic trap with variable stiffness. The particle position $x$ obeys the overdamped Langevin equation,
\begin{align}
\gamma\dot{x} = -kx +\sqrt{2\gamma k_{\rm B}T} \, \eta,
\end{align}
in a trap of time-dependent strength $k$, damping coefficient $\gamma$, environmental temperature $T$, Boltzmann's constant $k_{\rm B}$, and Gaussian white noise $\eta$. The control parameter is the trap strength $\lambda = k$, so the conjugate force is $f \equiv -\partial U/\partial k = -\frac{1}{2}x^{2}$. The joint probability distribution of the particle position and work obeys~\cite{Imparato2007,Solon2018}
\begin{align}
\gamma\frac{\partial p(x,w,t)}{\partial t} = -\frac{1}{2}\gamma\dot{k}x^2\frac{\partial p}{\partial w} + \frac{\partial }{\partial x}\left(kxp+\frac{1}{\beta}\frac{\partial p}{\partial x}\right) \ .
\label{Joint position and work distribution}
\end{align}
From this, we can solve for any moment of either the work or position distribution, subject to an initial equilibrium condition $p(x,w,t=0) = \pi(x|k_{\rm i})\delta(w)= [2\pi/(\beta k_{\rm i})]^{-1/2}\exp\{-\beta k_{\rm i}x^2/2\}\delta(w)$, with $\delta(w)$ the Dirac delta function. 

\begin{figure} 
	\includegraphics[width=0.5\textwidth]{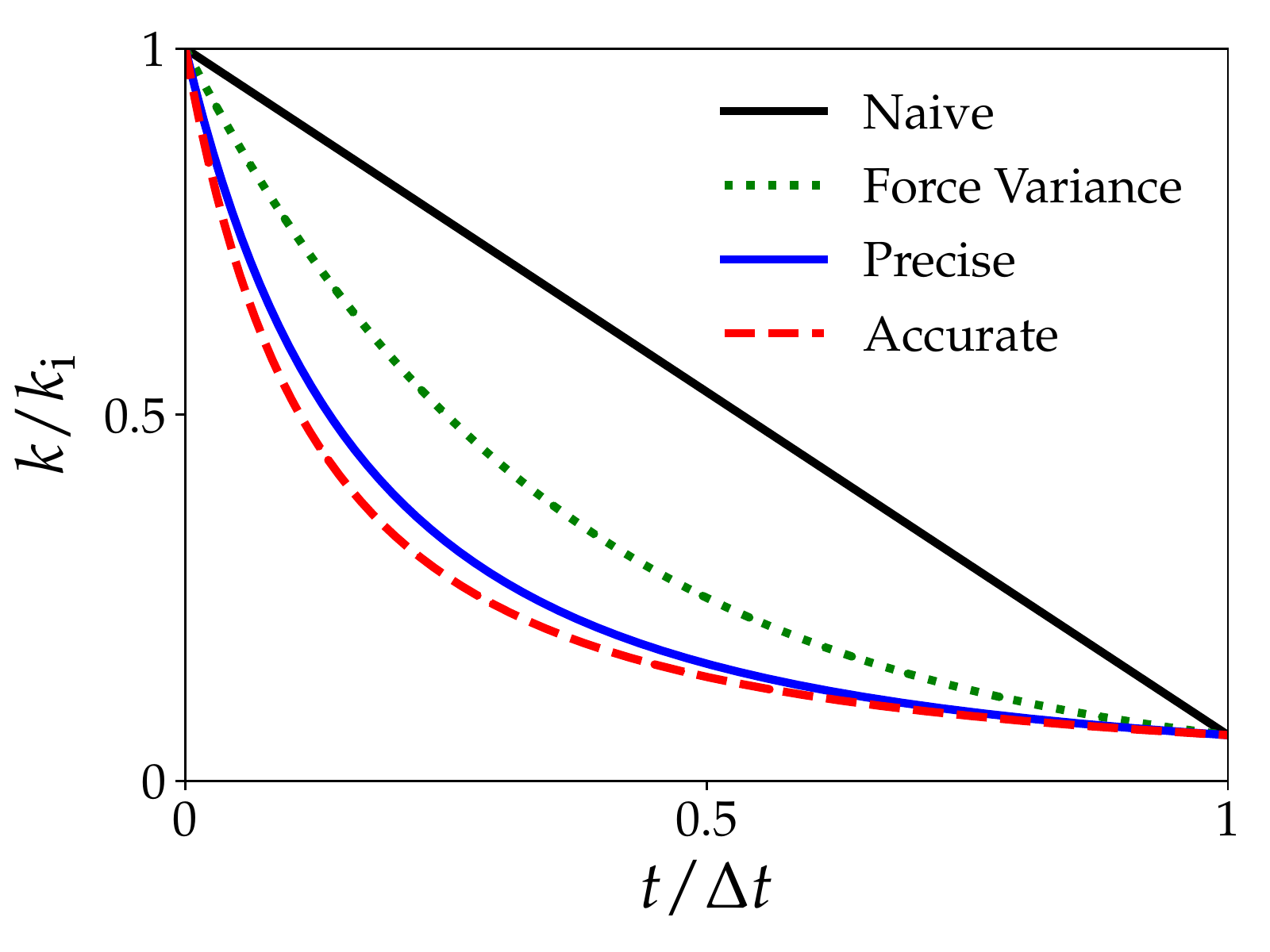}
	\caption{The control parameter (spring constant $k$) as a function of time for three different protocols in the model system of a breathing harmonic trap. The control parameter drives the system between relatively strongly and weakly confined states ($k_{\rm f}/k_{\rm i} =1/16$). `Naive' denotes the constant-velocity protocol,	`Force Variance' the force-variance-optimized protocol proceeding according to $\dot{k}\propto k$, `Accurate' the minimum-variance protocol with $\dot{k}\propto k^{3/2}$, and `Precise' the minimum-bias protocol with $\dot{k}\propto k^{5/3}$.}
	\label{Protocol}
\end{figure}

The force variance~\eqref{force variance}, the Stokes'~\eqref{rank 2 friction} and the supra-Stokes' coefficients~\eqref{rank 3 friction}, and the third integral covariance function~\eqref{rank n friction} are 
\begin{subequations}
	\begin{align}
	\langle \delta f^{2}\rangle_{k(t)} &= \frac{1}{2k^2} \\
	\zeta[k(t)] &= \frac{\beta\gamma}{4k^3} \\
	\zeta^{(2)}[k(t)] &=  \frac{\beta^2\gamma^2}{2k^5} \\
	\mathcal{C}^{(3)}[k(t),t] &\approx \beta^2\gamma\left(\frac{3t}{16k^6}+ \frac{51\gamma}{32k^7}\right) \ ,
	\label{HT 3rd integral covariance}
	\end{align}
\end{subequations} 
respectively. In \eqref{HT 3rd integral covariance}, we neglect terms of order $t^{-n}$ for $n \ge 1$. Since we only consider one-dimensional control, we can analytically determine the minimum-variance (precise) and minimum-bias (accurate) protocols that minimize Eq.~\eqref{BAR Variance Friction} and \eqref{BAR Bias Friction}, respectively. The force-variance-optimized protocol proceeds at velocity $\dot{k}\propto k$, minimum-variance protocol at $\dot{k} \propto k^{3/2}$, and minimum-bias protocol at $\dot{k} \propto k^{5/3}$. Figure~\ref{Protocol} plots these protocols alongside the naive (constant-velocity) protocol.

\begin{figure}
	\includegraphics[width=0.5\textwidth]{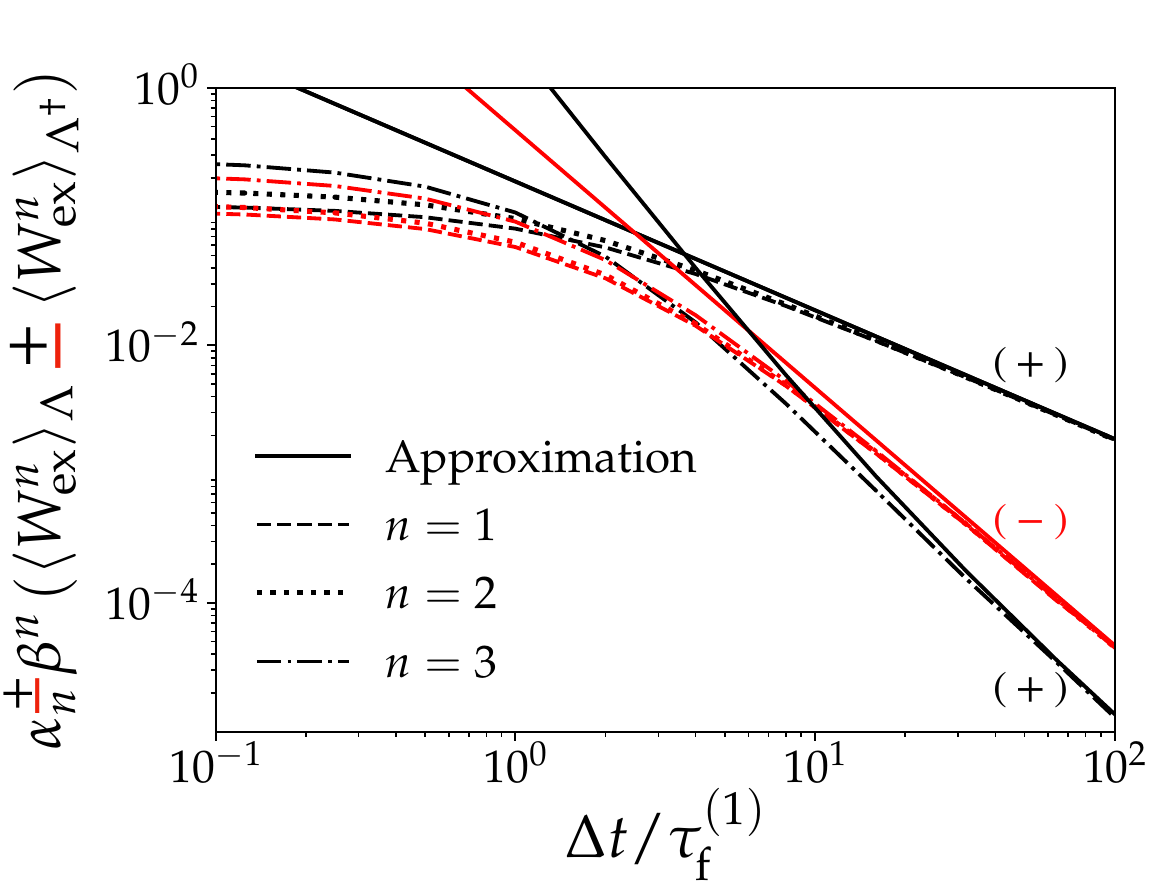}
	\caption{The sum (black, $+$) and the difference (red, $-$) of the moments of excess work for forward and reverse protocols of the breathing harmonic trap, as a function of protocol duration $\Delta t$ (scaled by the slowest relaxation time $\tau^{(1)}_{\rm f} = \gamma/(2k_{\rm f})$). Solid lines are the near-equilibrium approximations, given by Eqs.~\eqref{Mean_work}, \eqref{Variance friction approximation}, and \eqref{Higher order moment friction approximation}. Dashed ($n=1$), dotted ($n=2$), and dash-dotted ($n=3$) curves show exact results. The protocol $k(t)$ is linear with $k_{\rm f}/k_{\rm i} = 1/2$. The coefficients $\alpha^{+}_{1}=1/2$, $\alpha^{-}_{1}=2$, $\alpha^{+}_{2}=1/4$, $\alpha^{-}_{2}=1/3$, $\alpha^{+}_{3}=1/2$ and $\alpha^{-}_{3}=1/4$ are chosen such that any moment approximated by the same friction coefficient collapses onto a single curve. In all cases the exact calculations agree with the approximation for large $\Delta t/\tau^{(1)}_{\rm f}$.}
	\label{Sum_Diff_Work}
\end{figure}

As a simple test of the approximations of Eqs.~\eqref{Mean_work}, \eqref{Variance friction approximation}, and \eqref{Higher order moment friction approximation}, Fig.~\ref{Sum_Diff_Work} plots the sum and difference of the first three moments of the excess work from forward and reverse protocols. We choose the forward protocol to be a decrease in the control parameter (decreasing $k$), and therefore the reverse increases the control parameter (increasing $k$). In all cases the exact calculations agree with the approximation for slow protocols (large $\Delta t/\tau^{(1)}_{\rm f}$ for slowest relaxation time $\tau^{(1)}_{\rm f} = \gamma/(2k_{\rm f})$), and overestimate for fast protocols (small $\Delta t/\tau^{(1)}_{\rm f}$). For large $\Delta t/\tau^{(1)}_{\rm f}$, the first and second moments are approximated by \eqref{Mean_work} and \eqref{Variance friction approximation} so the sums ($n=1$ and $2$ ($+$)) are both approximated by the Stokes' friction~\eqref{rank 2 friction} and decrease as $1/\Delta t$,  while the differences ($n=1$ and $2$ ($-$)) are proportional to the supra-Stokes' tensor~\eqref{rank 3 friction} and decrease as $1/(\Delta t)^{2}$. The third moment is approximated by~\eqref{Higher order moment friction approximation}, so the sum ($n=3$ ($+$)) is approximated by the third integral covariance~\eqref{rank n friction} and decays as $1/(\Delta t)^2$ (due to the term linear in $t$ in~\eqref{HT 3rd integral covariance}) while the difference ($n=3$ ($-$)) is proportional to the supra-Stokes tensor and decays as $1/(\Delta t)^2$.

\section{Finite Integration Bounds}
\label{Finite Bounds}

The infinite integration bound on the friction assumes that correlations relax quickly relative to the protocol speed. 
For the first two moments, despite the finite integration bound generally yielding a more accurate approximation, the infinite bound considerably simplifies the approximation and allows for straightforward protocol optimization. 

The effect of finite integration bounds is significant for two calculations: approximation of fourth- and higher-order moments of the excess work, and next-order approximations (including both leading- and next-to-leading-order contributions) for the moments. In the former, one must treat the bound as finite since the $n$-time covariance functions do not decay to zero for some subspace of large time arguments.

In more detail, consider the four-time covariance (kurtosis) $\left\langle \delta f_{i}(0)\delta f_{j}(t_{2})\delta f_{k}(t_{3})\delta f_{\ell}(t_{4})\right\rangle_{\boldsymbol{\lambda}(t)}$. When any one of the four times $\{0, t_2, t_3, t_4\}$ significantly differs from the others, the conjugate forces decorrelate, and the kurtosis decays to zero; however, when $t_{3}\sim t_{4} \gg t_{2} \sim 0$, any variables separated by significant time decorrelate, and the kurtosis approaches $\langle \delta f_{i}(0)\delta f_{j}(t_{2})\rangle_{\boldsymbol{\lambda} (t)}\langle \delta f_{k}(t_{3})\delta f_{\ell}(t_{4})\rangle_{\boldsymbol{\lambda} (t)}$. This limit represents a plane in the $(t_{2},t_{3},t_{4})$ parameter space where the kurtosis asymptotes to a finite value even for large time arguments. Integrating the above and multiplying by three (to account for permutations of the indices $1,2,3,4$) yields the integral four-time covariance in the limit $t_{3}\sim t_{4} \gg t_{2} \sim 0$,
\begin{subequations}
	\begin{align}
	\mathcal{C}_{ijk\ell}^{(3)}[\boldsymbol{\lambda}(t),t]&=3t \, \mathcal{C}^{(1)}_{ij}[\boldsymbol{\lambda}(t),t]\mathcal{C}^{(1)}_{k\ell}[\boldsymbol{\lambda}(t),t]  \\
	& = 3t \, \zeta^{(1)}_{ij}[\boldsymbol{\lambda}(t)]\zeta^{(1)}_{k\ell}[\boldsymbol{\lambda}(t)] \ , \quad t\to\infty \ .
	\end{align}
\end{subequations}
Since this is the only case that remains finite as $t\to\infty$, the second line is the approximation (to highest order in $t$) for the integral four-time covariance. Parallel arguments hold for the higher-order moments.

When approximating the average excess work~\eqref{total excess work} with both the Stokes' and supra-Stokes' tensors~\eqref{total friction}, finite integration bounds on the Stokes' friction may be necessary. For finite integration bounds on the Stokes' friction we replace \eqref{total friction} with 
\begin{align}
\mathcal{C}_{ij}[\boldsymbol{\lambda}(t),\dot{\lambda}(t),t] \equiv \mathcal{C}^{(1)}_{ij}[\boldsymbol{\lambda}(t),t]+\frac{1}{4}\zeta^{(2)}_{ijk}[\boldsymbol{\lambda}(t)]\dot{\lambda}_{k}(t) \ .
\label{total integral covariance}
\end{align}
For systems with weakly skewed conjugate-force fluctuation, the contribution from the finite bound on the first term can be comparable in magnitude to the contribution from the supra-Stokes' tensor. 

\begin{figure}
	\includegraphics[width=0.5\textwidth]{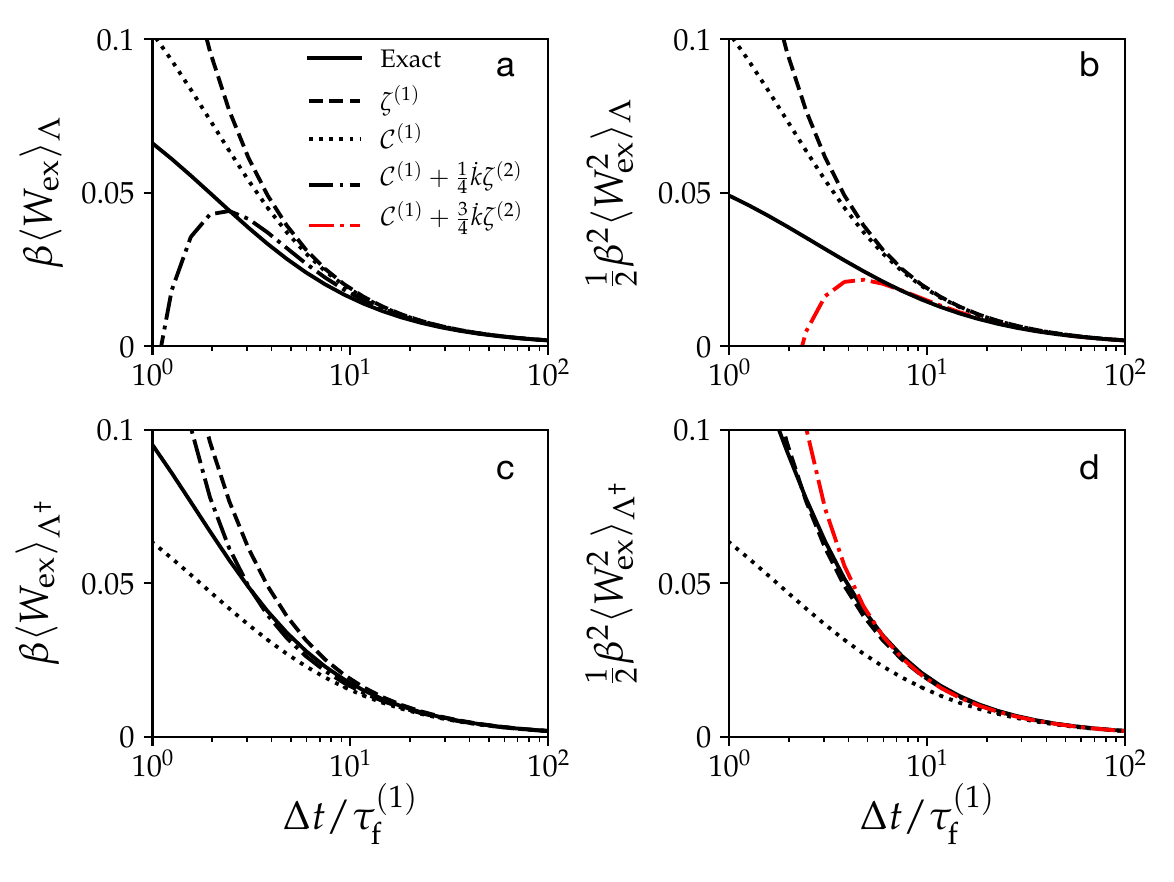}
	\caption{The first and second moments of excess works for forward (a,b) and reverse (c,d) protocols of the breathing harmonic trap as a function of protocol time $\Delta t$ (scaled by the slowest relaxation time $\tau^{(1)}_{\rm f} = \gamma/(2k_{\rm f})$). The dashed, dotted, and dash-dotted lines are different forms of the near-equilibrium approximation. The protocol $k(t)$ is linear with $k_{\rm f}/k_{\rm i} = 1/2$.}
	\label{Finite_Forward_Reverse}
\end{figure}

Figures~\ref{Finite_Forward_Reverse} shows the first and second moments of the excess work for forward and reverse protocols, compared to different approximations. For the mean excess work, the supra-Stokes' tensor contributes with opposite sign from the Stokes' friction for forward (decreasing $k$) protocols and with same sign for reverse (increasing $k$). Since the Stokes'-friction approximation overestimates the excess work in both cases (a,c), adding the supra-Stokes' tensor reduces the accuracy of the approximation for the reverse excess work. This effect is an artifact of the infinite integral bound; indeed, if the bound is kept finite ($\mathcal{C}^{(1)}$ rather than $\zeta^{(1)}$), the first integral autocovariance overestimates for forward protocols and underestimates for reverse protocols, and adding the supra-Stokes' tensor improves the approximation in both cases. Finally, we note that the Stokes' friction does not always overestimate for both forward and reverse protocols. For the second moment in (b,d), the Stokes' friction overestimates for forward protocols and underestimates for reverse protocols, and the supra-Stokes' tensor improves the approximation when either the Stokes' friction or first integral covariance are used.

\section{Finsler Geometry}\label{Finsler Geometry}

The addition of the supra-Stokes' tensor~\eqref{rank 3 friction} comes at the cost of a more complex geometric structure. In contrast to Riemannian geometry, Finsler geometry is not restricted to a quadratic norm. In general, the inner products are not characterized by points, but rather by points and directions. Despite this, several useful concepts from Riemannian geometry (notably curvature, length, and geodesics) generalize. There are therefore standard procedures to find the geodesics despite the more complex landscapes induced by the generalized cubic Finsler metric~\eqref{total friction}.

Finsler geometry has several applications in both physics and biology.~\cite{Antonelli2013} In the thermodynamic context, the \emph{flexion} tensor~\cite{Sellentin2014}
\begin{align}
\mathcal{F}_{ijk}[\boldsymbol{\lambda}(t)]   \equiv \left\langle\frac{\partial^3 \ln \pi (x|\boldsymbol{\lambda})}{\partial \lambda_{i}\partial\lambda_{j}\partial\lambda_{k}} \right\rangle_{\boldsymbol{\lambda} (t)} 
\end{align}
is a Finsler metric arising as the third-order contribution to the near-equilibrium expansion of the relative entropy (Kullback-Leibler divergence) of a probability distribution relative to the equilibrium distribution:~\cite{Antonelli2013}
\begin{subequations}
	\begin{align}
	D(\pi(x|\boldsymbol{\lambda})||\pi(x'|\boldsymbol{\lambda})|) &\equiv \int_{-\infty}^{\infty}\md x~ \pi(x|\boldsymbol{\lambda})\ln\frac{\pi(x'|\boldsymbol{\lambda})}{\pi(x|\boldsymbol{\lambda})} \\
	& \approx \mathcal{I}_{ij}\Delta\lambda_{i}\Delta\lambda_{j} +\mathcal{F}_{ijk}\Delta\lambda_{i}\Delta\lambda_{j}\Delta\lambda_{k} \ .
	\end{align}
\end{subequations}
The coskewness tensor~\eqref{coskewness} is related to the flexion tensor by
\begin{align}
\mathcal{F}_{ijk}[\lambda(t)]  = &\langle\delta f_{i}\delta f_{j}\delta f_{k}\rangle_{\boldsymbol{\lambda}(t)}
\\
&+\sum_{\sigma_{ijk}}\left\langle\frac{\partial^2 \ln \pi (x|\lambda)}{\partial \lambda_{i}\partial\lambda_{{j}}} \frac{\partial \ln \pi (x|\lambda)}{\partial \lambda_{{k}}}\right\rangle_{\lambda (t)} \ ,\nonumber
\end{align}
where the sum is over permutations of the indices $\sigma_{ijk} = \{123,132,321\} $. In weak gravitational lensing~\cite{Goldberg2005,Bacon2006} the flexion tensor is the third-order correction to the shapes of images, which is described as flexing the shape of images from an ellipse towards a banana shape. Similarly, the supra-Stokes' tensor (product of coskewness tensor and integral double relaxation time~\eqref{coskewness}) skews the thermodynamic geometry of minimum-dissipation protocols, as demonstrated in Fig.~\ref{Friction_Metric}.


\begin{thebibliography}{58}
	\expandafter\ifx\csname natexlab\endcsname\relax\def\natexlab#1{#1}\fi
	\expandafter\ifx\csname bibnamefont\endcsname\relax
	\def\bibnamefont#1{#1}\fi
	\expandafter\ifx\csname bibfnamefont\endcsname\relax
	\def\bibfnamefont#1{#1}\fi
	\expandafter\ifx\csname citenamefont\endcsname\relax
	\def\citenamefont#1{#1}\fi
	\expandafter\ifx\csname url\endcsname\relax
	\def\url#1{\texttt{#1}}\fi
	\expandafter\ifx\csname urlprefix\endcsname\relax\def\urlprefix{URL }\fi
	\providecommand{\bibinfo}[2]{#2}
	\providecommand{\eprint}[2][]{\url{#2}}
	
	\bibitem[{\citenamefont{Gapsys et~al.}(2015)\citenamefont{Gapsys, Michielssens,
			Peters, de~Groot, and Leonov}}]{Gapsys2015}
	\bibinfo{author}{\bibfnamefont{V.}~\bibnamefont{Gapsys}},
	\bibinfo{author}{\bibfnamefont{S.}~\bibnamefont{Michielssens}},
	\bibinfo{author}{\bibfnamefont{J.~H.} \bibnamefont{Peters}},
	\bibinfo{author}{\bibfnamefont{B.~L.} \bibnamefont{de~Groot}},
	\bibnamefont{and} \bibinfo{author}{\bibfnamefont{H.}~\bibnamefont{Leonov}},
	\emph{\bibinfo{title}{Calculation of Binding Free Energies}}
	(\bibinfo{publisher}{Springer New York}, \bibinfo{address}{New York, NY},
	\bibinfo{year}{2015}), pp. \bibinfo{pages}{173--209}.
	
	\bibitem[{\citenamefont{Schindler et~al.}(2020)\citenamefont{Schindler,
			Baumann, Blum, B{\"o}se, Buchstaller, Burgdorf, Cappel, Chekler, Czodrowski,
			Dorsch et~al.}}]{Schindler2020}
	\bibinfo{author}{\bibfnamefont{C.~E.} \bibnamefont{Schindler}},
	\bibinfo{author}{\bibfnamefont{H.}~\bibnamefont{Baumann}},
	\bibinfo{author}{\bibfnamefont{A.}~\bibnamefont{Blum}},
	\bibinfo{author}{\bibfnamefont{D.}~\bibnamefont{B{\"o}se}},
	\bibinfo{author}{\bibfnamefont{H.-P.} \bibnamefont{Buchstaller}},
	\bibinfo{author}{\bibfnamefont{L.}~\bibnamefont{Burgdorf}},
	\bibinfo{author}{\bibfnamefont{D.}~\bibnamefont{Cappel}},
	\bibinfo{author}{\bibfnamefont{E.}~\bibnamefont{Chekler}},
	\bibinfo{author}{\bibfnamefont{P.}~\bibnamefont{Czodrowski}},
	\bibinfo{author}{\bibfnamefont{D.}~\bibnamefont{Dorsch}},
	\bibnamefont{et~al.}, \bibinfo{journal}{J. Chem. Inf. Model}
	\textbf{\bibinfo{volume}{60}}, \bibinfo{pages}{5457} (\bibinfo{year}{2020}).
	
	\bibitem[{\citenamefont{Kuhn et~al.}(2017)\citenamefont{Kuhn, Tich\'{y}, Wang,
			Robinson, Martin, Kuglstatter, Benz, Giroud, Schirmeister, Abel
			et~al.}}]{Kuhn2017}
	\bibinfo{author}{\bibfnamefont{B.}~\bibnamefont{Kuhn}},
	\bibinfo{author}{\bibfnamefont{M.}~\bibnamefont{Tich\'{y}}},
	\bibinfo{author}{\bibfnamefont{L.}~\bibnamefont{Wang}},
	\bibinfo{author}{\bibfnamefont{S.}~\bibnamefont{Robinson}},
	\bibinfo{author}{\bibfnamefont{R.~E.} \bibnamefont{Martin}},
	\bibinfo{author}{\bibfnamefont{A.}~\bibnamefont{Kuglstatter}},
	\bibinfo{author}{\bibfnamefont{J.}~\bibnamefont{Benz}},
	\bibinfo{author}{\bibfnamefont{M.}~\bibnamefont{Giroud}},
	\bibinfo{author}{\bibfnamefont{T.}~\bibnamefont{Schirmeister}},
	\bibinfo{author}{\bibfnamefont{R.}~\bibnamefont{Abel}}, \bibnamefont{et~al.},
	\bibinfo{journal}{J. Med. Chem} \textbf{\bibinfo{volume}{60}},
	\bibinfo{pages}{2485} (\bibinfo{year}{2017}).
	
	\bibitem[{\citenamefont{Ciordia et~al.}(2016)\citenamefont{Ciordia,
			P{\'e}rez-Benito, Delgado, Trabanco, and Tresadern}}]{Ciordia2016}
	\bibinfo{author}{\bibfnamefont{M.}~\bibnamefont{Ciordia}},
	\bibinfo{author}{\bibfnamefont{L.}~\bibnamefont{P{\'e}rez-Benito}},
	\bibinfo{author}{\bibfnamefont{F.}~\bibnamefont{Delgado}},
	\bibinfo{author}{\bibfnamefont{A.~A.} \bibnamefont{Trabanco}},
	\bibnamefont{and}
	\bibinfo{author}{\bibfnamefont{G.}~\bibnamefont{Tresadern}},
	\bibinfo{journal}{J. Chem. Inf. Model} \textbf{\bibinfo{volume}{56}},
	\bibinfo{pages}{1856} (\bibinfo{year}{2016}).
	
	\bibitem[{\citenamefont{Wang et~al.}(2015)\citenamefont{Wang, Wu, Deng, Kim,
			Pierce, Krilov, Lupyan, Robinson, Dahlgren, Greenwood et~al.}}]{Wang2015}
	\bibinfo{author}{\bibfnamefont{L.}~\bibnamefont{Wang}},
	\bibinfo{author}{\bibfnamefont{Y.}~\bibnamefont{Wu}},
	\bibinfo{author}{\bibfnamefont{Y.}~\bibnamefont{Deng}},
	\bibinfo{author}{\bibfnamefont{B.}~\bibnamefont{Kim}},
	\bibinfo{author}{\bibfnamefont{L.}~\bibnamefont{Pierce}},
	\bibinfo{author}{\bibfnamefont{G.}~\bibnamefont{Krilov}},
	\bibinfo{author}{\bibfnamefont{D.}~\bibnamefont{Lupyan}},
	\bibinfo{author}{\bibfnamefont{S.}~\bibnamefont{Robinson}},
	\bibinfo{author}{\bibfnamefont{M.~K.} \bibnamefont{Dahlgren}},
	\bibinfo{author}{\bibfnamefont{J.}~\bibnamefont{Greenwood}},
	\bibnamefont{et~al.}, \bibinfo{journal}{J. Am. Chem. Soc.}
	\textbf{\bibinfo{volume}{137}}, \bibinfo{pages}{2695} (\bibinfo{year}{2015}).
	
	\bibitem[{\citenamefont{Aldeghi et~al.}(2018)\citenamefont{Aldeghi, Gapsys, and
			de~Groot}}]{Aldeghi2018}
	\bibinfo{author}{\bibfnamefont{M.}~\bibnamefont{Aldeghi}},
	\bibinfo{author}{\bibfnamefont{V.}~\bibnamefont{Gapsys}}, \bibnamefont{and}
	\bibinfo{author}{\bibfnamefont{B.~L.} \bibnamefont{de~Groot}},
	\bibinfo{journal}{ACS Cent. Sci.} \textbf{\bibinfo{volume}{4}},
	\bibinfo{pages}{1708} (\bibinfo{year}{2018}).
	
	\bibitem[{\citenamefont{Chodera et~al.}(2011)\citenamefont{Chodera, Mobley,
			Shirts, Dixon, Branson, and Pande}}]{Chodera2011}
	\bibinfo{author}{\bibfnamefont{J.~D.} \bibnamefont{Chodera}},
	\bibinfo{author}{\bibfnamefont{D.~L.} \bibnamefont{Mobley}},
	\bibinfo{author}{\bibfnamefont{M.~R.} \bibnamefont{Shirts}},
	\bibinfo{author}{\bibfnamefont{R.~W.} \bibnamefont{Dixon}},
	\bibinfo{author}{\bibfnamefont{K.}~\bibnamefont{Branson}}, \bibnamefont{and}
	\bibinfo{author}{\bibfnamefont{V.~S.} \bibnamefont{Pande}},
	\bibinfo{journal}{Curr. Opin. Struc. Biol.} \textbf{\bibinfo{volume}{21}},
	\bibinfo{pages}{150} (\bibinfo{year}{2011}).
	
	\bibitem[{\citenamefont{Pohorille et~al.}(2010)\citenamefont{Pohorille,
			Jarzynski, and Chipot}}]{Pohorille2010}
	\bibinfo{author}{\bibfnamefont{A.}~\bibnamefont{Pohorille}},
	\bibinfo{author}{\bibfnamefont{C.}~\bibnamefont{Jarzynski}},
	\bibnamefont{and} \bibinfo{author}{\bibfnamefont{C.}~\bibnamefont{Chipot}},
	\bibinfo{journal}{J. Phys. Chem. B} \textbf{\bibinfo{volume}{114}},
	\bibinfo{pages}{10235} (\bibinfo{year}{2010}).
	
	\bibitem[{\citenamefont{Gore et~al.}(2003)\citenamefont{Gore, Ritort, and
			Bustamante}}]{Gore2003}
	\bibinfo{author}{\bibfnamefont{J.}~\bibnamefont{Gore}},
	\bibinfo{author}{\bibfnamefont{F.}~\bibnamefont{Ritort}}, \bibnamefont{and}
	\bibinfo{author}{\bibfnamefont{C.}~\bibnamefont{Bustamante}},
	\bibinfo{journal}{Proc. Natl. Acad. Sci. U.S.A}
	\textbf{\bibinfo{volume}{100}}, \bibinfo{pages}{12564}
	(\bibinfo{year}{2003}).
	
	\bibitem[{\citenamefont{Jarzynski}(1997)}]{Jarzynski1997}
	\bibinfo{author}{\bibfnamefont{C.}~\bibnamefont{Jarzynski}},
	\bibinfo{journal}{Phys. Rev. Lett.} \textbf{\bibinfo{volume}{78}},
	\bibinfo{pages}{2690} (\bibinfo{year}{1997}).
	
	\bibitem[{\citenamefont{Bennett}(1976)}]{Bennett1976}
	\bibinfo{author}{\bibfnamefont{C.~H.} \bibnamefont{Bennett}},
	\bibinfo{journal}{J. Comput. Phys.} \textbf{\bibinfo{volume}{22}},
	\bibinfo{pages}{245} (\bibinfo{year}{1976}).
	
	\bibitem[{\citenamefont{Shirts et~al.}(2003)\citenamefont{Shirts, Bair, Hooker,
			and Pande}}]{Shirts2003}
	\bibinfo{author}{\bibfnamefont{M.~R.} \bibnamefont{Shirts}},
	\bibinfo{author}{\bibfnamefont{E.}~\bibnamefont{Bair}},
	\bibinfo{author}{\bibfnamefont{G.}~\bibnamefont{Hooker}}, \bibnamefont{and}
	\bibinfo{author}{\bibfnamefont{V.~S.} \bibnamefont{Pande}},
	\bibinfo{journal}{Phys. Rev. Lett.} \textbf{\bibinfo{volume}{91}},
	\bibinfo{pages}{140601} (\bibinfo{year}{2003}).
	
	\bibitem[{\citenamefont{Maragakis et~al.}(2006)\citenamefont{Maragakis,
			Spichty, and Karplus}}]{Maragakis2006}
	\bibinfo{author}{\bibfnamefont{P.}~\bibnamefont{Maragakis}},
	\bibinfo{author}{\bibfnamefont{M.}~\bibnamefont{Spichty}}, \bibnamefont{and}
	\bibinfo{author}{\bibfnamefont{M.}~\bibnamefont{Karplus}},
	\bibinfo{journal}{Phys. Rev. Lett.} \textbf{\bibinfo{volume}{96}},
	\bibinfo{pages}{100602} (\bibinfo{year}{2006}).
	
	\bibitem[{\citenamefont{Reinhardt and Hunter~III}(1992)}]{Reinhardt1992}
	\bibinfo{author}{\bibfnamefont{W.~P.} \bibnamefont{Reinhardt}}
	\bibnamefont{and} \bibinfo{author}{\bibfnamefont{J.~E.}
		\bibnamefont{Hunter~III}}, \bibinfo{journal}{J. Chem. Phys.}
	\textbf{\bibinfo{volume}{97}}, \bibinfo{pages}{1599} (\bibinfo{year}{1992}).
	
	\bibitem[{\citenamefont{Hunter~III et~al.}(1993)\citenamefont{Hunter~III,
			Reinhardt, and Davis}}]{Hunter1993}
	\bibinfo{author}{\bibfnamefont{J.~E.} \bibnamefont{Hunter~III}},
	\bibinfo{author}{\bibfnamefont{W.~P.} \bibnamefont{Reinhardt}},
	\bibnamefont{and} \bibinfo{author}{\bibfnamefont{T.~F.} \bibnamefont{Davis}},
	\bibinfo{journal}{J. Chem. Phys.} \textbf{\bibinfo{volume}{99}},
	\bibinfo{pages}{6856} (\bibinfo{year}{1993}).
	
	\bibitem[{\citenamefont{Jarque and Tidor}(1997)}]{Jarque1997}
	\bibinfo{author}{\bibfnamefont{C.}~\bibnamefont{Jarque}} \bibnamefont{and}
	\bibinfo{author}{\bibfnamefont{B.}~\bibnamefont{Tidor}}, \bibinfo{journal}{J.
		Phys. Chem. B} \textbf{\bibinfo{volume}{101}}, \bibinfo{pages}{9402}
	(\bibinfo{year}{1997}).
	
	\bibitem[{\citenamefont{Weinhold}(1975)}]{Weinhold1975}
	\bibinfo{author}{\bibfnamefont{F.}~\bibnamefont{Weinhold}},
	\bibinfo{journal}{J. Chem. Phys.} \textbf{\bibinfo{volume}{63}},
	\bibinfo{pages}{2479} (\bibinfo{year}{1975}).
	
	\bibitem[{\citenamefont{Salamon and Berry}(1983)}]{Salamon1983}
	\bibinfo{author}{\bibfnamefont{P.}~\bibnamefont{Salamon}} \bibnamefont{and}
	\bibinfo{author}{\bibfnamefont{R.~S.} \bibnamefont{Berry}},
	\bibinfo{journal}{Phys. Rev. Lett.} \textbf{\bibinfo{volume}{51}},
	\bibinfo{pages}{1127} (\bibinfo{year}{1983}).
	
	\bibitem[{\citenamefont{Sch{\"o}n}(1996)}]{Schon1996}
	\bibinfo{author}{\bibfnamefont{J.~C.} \bibnamefont{Sch{\"o}n}},
	\bibinfo{journal}{J. Chem. Phys.} \textbf{\bibinfo{volume}{105}},
	\bibinfo{pages}{10072} (\bibinfo{year}{1996}).
	
	\bibitem[{\citenamefont{Miller and Reinhardt}(2000)}]{Miller2000}
	\bibinfo{author}{\bibfnamefont{M.~A.} \bibnamefont{Miller}} \bibnamefont{and}
	\bibinfo{author}{\bibfnamefont{W.~P.} \bibnamefont{Reinhardt}},
	\bibinfo{journal}{J. Chem. Phys.} \textbf{\bibinfo{volume}{113}},
	\bibinfo{pages}{7035} (\bibinfo{year}{2000}).
	
	\bibitem[{\citenamefont{Crooks}(2007)}]{Crooks2007}
	\bibinfo{author}{\bibfnamefont{G.~E.} \bibnamefont{Crooks}},
	\bibinfo{journal}{Phys. Rev. Lett.} \textbf{\bibinfo{volume}{99}},
	\bibinfo{pages}{100602} (\bibinfo{year}{2007}).
	
	\bibitem[{\citenamefont{Shenfeld et~al.}(2009)\citenamefont{Shenfeld, Xu,
			Eastwood, Dror, and Shaw}}]{Shenfeld2009}
	\bibinfo{author}{\bibfnamefont{D.~K.} \bibnamefont{Shenfeld}},
	\bibinfo{author}{\bibfnamefont{H.}~\bibnamefont{Xu}},
	\bibinfo{author}{\bibfnamefont{M.~P.} \bibnamefont{Eastwood}},
	\bibinfo{author}{\bibfnamefont{R.~O.} \bibnamefont{Dror}}, \bibnamefont{and}
	\bibinfo{author}{\bibfnamefont{D.~E.} \bibnamefont{Shaw}},
	\bibinfo{journal}{Phys. Rev. E} \textbf{\bibinfo{volume}{80}},
	\bibinfo{pages}{046705} (\bibinfo{year}{2009}).
	
	\bibitem[{\citenamefont{Minh}(2020)}]{Minh2019}
	\bibinfo{author}{\bibfnamefont{D.~D.~L.} \bibnamefont{Minh}},
	\bibinfo{journal}{J. of Comp. Chem.} \textbf{\bibinfo{volume}{41}},
	\bibinfo{pages}{715} (\bibinfo{year}{2020}).
	
	\bibitem[{\citenamefont{Pham and Shirts}(2011)}]{Pham2011}
	\bibinfo{author}{\bibfnamefont{T.~T.} \bibnamefont{Pham}} \bibnamefont{and}
	\bibinfo{author}{\bibfnamefont{M.~R.} \bibnamefont{Shirts}},
	\bibinfo{journal}{J. Chem. Phys.} \textbf{\bibinfo{volume}{135}},
	\bibinfo{pages}{034114} (\bibinfo{year}{2011}).
	
	\bibitem[{\citenamefont{Pham and Shirts}(2012)}]{Pham2012}
	\bibinfo{author}{\bibfnamefont{T.~T.} \bibnamefont{Pham}} \bibnamefont{and}
	\bibinfo{author}{\bibfnamefont{M.~R.} \bibnamefont{Shirts}},
	\bibinfo{journal}{J. Chem. Phys.} \textbf{\bibinfo{volume}{136}},
	\bibinfo{pages}{124120} (\bibinfo{year}{2012}).
	
	\bibitem[{\citenamefont{Park and Im}(2014)}]{Park2014}
	\bibinfo{author}{\bibfnamefont{S.}~\bibnamefont{Park}} \bibnamefont{and}
	\bibinfo{author}{\bibfnamefont{W.}~\bibnamefont{Im}}, \bibinfo{journal}{J.
		Chem. Theory Comput.} \textbf{\bibinfo{volume}{10}}, \bibinfo{pages}{2719}
	(\bibinfo{year}{2014}).
	
	\bibitem[{\citenamefont{Sivak and Crooks}(2012)}]{OptimalPaths}
	\bibinfo{author}{\bibfnamefont{D.~A.} \bibnamefont{Sivak}} \bibnamefont{and}
	\bibinfo{author}{\bibfnamefont{G.~E.} \bibnamefont{Crooks}},
	\bibinfo{journal}{Phys. Rev. Lett.} \textbf{\bibinfo{volume}{108}},
	\bibinfo{pages}{190602} (\bibinfo{year}{2012}).
	
	\bibitem[{\citenamefont{Deffner and Bonan{\c{c}}a}(2020)}]{Deffner2020}
	\bibinfo{author}{\bibfnamefont{S.}~\bibnamefont{Deffner}} \bibnamefont{and}
	\bibinfo{author}{\bibfnamefont{M.~V.~S.} \bibnamefont{Bonan{\c{c}}a}},
	\bibinfo{journal}{EPL} \textbf{\bibinfo{volume}{131}}, \bibinfo{pages}{20001}
	(\bibinfo{year}{2020}).
	
	\bibitem[{\citenamefont{Kim et~al.}(2012)\citenamefont{Kim, Kim, Talkner, and
			Yi}}]{Kim2012}
	\bibinfo{author}{\bibfnamefont{S.}~\bibnamefont{Kim}},
	\bibinfo{author}{\bibfnamefont{Y.~W.} \bibnamefont{Kim}},
	\bibinfo{author}{\bibfnamefont{P.}~\bibnamefont{Talkner}}, \bibnamefont{and}
	\bibinfo{author}{\bibfnamefont{J.}~\bibnamefont{Yi}}, \bibinfo{journal}{Phys.
		Rev. E} \textbf{\bibinfo{volume}{86}}, \bibinfo{pages}{041130}
	(\bibinfo{year}{2012}).
	
	\bibitem[{\citenamefont{Crooks}(1999)}]{Crooks1999}
	\bibinfo{author}{\bibfnamefont{G.~E.} \bibnamefont{Crooks}},
	\bibinfo{journal}{Phys. Rev. E} \textbf{\bibinfo{volume}{60}},
	\bibinfo{pages}{2721} (\bibinfo{year}{1999}).
	
	\bibitem[{\citenamefont{Crooks}(2000)}]{Crooks2000}
	\bibinfo{author}{\bibfnamefont{G.~E.} \bibnamefont{Crooks}},
	\bibinfo{journal}{Phys. Rev. E} \textbf{\bibinfo{volume}{61}},
	\bibinfo{pages}{2361} (\bibinfo{year}{2000}).
	
	\bibitem[{\citenamefont{Antonelli et~al.}(2013)\citenamefont{Antonelli,
			Ingarden, and Matsumoto}}]{Antonelli2013}
	\bibinfo{author}{\bibfnamefont{P.~L.} \bibnamefont{Antonelli}},
	\bibinfo{author}{\bibfnamefont{R.~S.} \bibnamefont{Ingarden}},
	\bibnamefont{and}
	\bibinfo{author}{\bibfnamefont{M.}~\bibnamefont{Matsumoto}},
	\emph{\bibinfo{title}{The theory of sprays and Finsler spaces with
			applications in physics and biology}}, vol.~\bibinfo{volume}{58}
	(\bibinfo{publisher}{Springer Science \& Business Media},
	\bibinfo{year}{2013}).
	
	\bibitem[{\citenamefont{Hirokawa et~al.}(2009)\citenamefont{Hirokawa, Noda,
			Tanaka, and Niwa}}]{Hirokawa2009}
	\bibinfo{author}{\bibfnamefont{N.}~\bibnamefont{Hirokawa}},
	\bibinfo{author}{\bibfnamefont{Y.}~\bibnamefont{Noda}},
	\bibinfo{author}{\bibfnamefont{Y.}~\bibnamefont{Tanaka}}, \bibnamefont{and}
	\bibinfo{author}{\bibfnamefont{S.}~\bibnamefont{Niwa}},
	\bibinfo{journal}{Nat. Rev. Mol. Cell Biol.} \textbf{\bibinfo{volume}{10}},
	\bibinfo{pages}{682} (\bibinfo{year}{2009}).
	
	\bibitem[{\citenamefont{Junge and Nelson}(2015)}]{Junge2015}
	\bibinfo{author}{\bibfnamefont{W.}~\bibnamefont{Junge}} \bibnamefont{and}
	\bibinfo{author}{\bibfnamefont{N.}~\bibnamefont{Nelson}},
	\bibinfo{journal}{Annu. Rev. Biochem.} \textbf{\bibinfo{volume}{84}},
	\bibinfo{pages}{631} (\bibinfo{year}{2015}).
	
	\bibitem[{\citenamefont{Schmiedl and Seifert}(2007)}]{Schmiedl2007}
	\bibinfo{author}{\bibfnamefont{T.}~\bibnamefont{Schmiedl}} \bibnamefont{and}
	\bibinfo{author}{\bibfnamefont{U.}~\bibnamefont{Seifert}},
	\bibinfo{journal}{Phys. Rev. Lett.} \textbf{\bibinfo{volume}{98}},
	\bibinfo{pages}{108301} (\bibinfo{year}{2007}).
	
	\bibitem[{\citenamefont{Gomez-Marin et~al.}(2008)\citenamefont{Gomez-Marin,
			Schmiedl, and Seifert}}]{Gomez2008}
	\bibinfo{author}{\bibfnamefont{A.}~\bibnamefont{Gomez-Marin}},
	\bibinfo{author}{\bibfnamefont{T.}~\bibnamefont{Schmiedl}}, \bibnamefont{and}
	\bibinfo{author}{\bibfnamefont{U.}~\bibnamefont{Seifert}},
	\bibinfo{journal}{J. Chem. Phys.} \textbf{\bibinfo{volume}{129}},
	\bibinfo{pages}{024114} (\bibinfo{year}{2008}).
	
	\bibitem[{\citenamefont{Geiger and Dellago}(2010)}]{Geiger2010}
	\bibinfo{author}{\bibfnamefont{P.}~\bibnamefont{Geiger}} \bibnamefont{and}
	\bibinfo{author}{\bibfnamefont{C.}~\bibnamefont{Dellago}},
	\bibinfo{journal}{Phys. Rev. E} \textbf{\bibinfo{volume}{81}},
	\bibinfo{pages}{021127} (\bibinfo{year}{2010}).
	
	\bibitem[{\citenamefont{Solon and Horowitz}(2018)}]{Solon2018}
	\bibinfo{author}{\bibfnamefont{A.~P.} \bibnamefont{Solon}} \bibnamefont{and}
	\bibinfo{author}{\bibfnamefont{J.~M.} \bibnamefont{Horowitz}},
	\bibinfo{journal}{Phys. Rev. Lett.} \textbf{\bibinfo{volume}{120}},
	\bibinfo{pages}{180605} (\bibinfo{year}{2018}).
	
	\bibitem[{\citenamefont{Pearlman and Kollman}(1989)}]{Pearlman1989}
	\bibinfo{author}{\bibfnamefont{D.~A.} \bibnamefont{Pearlman}} \bibnamefont{and}
	\bibinfo{author}{\bibfnamefont{P.~A.} \bibnamefont{Kollman}},
	\bibinfo{journal}{J. Chem. Phys} \textbf{\bibinfo{volume}{90}},
	\bibinfo{pages}{2460} (\bibinfo{year}{1989}).
	
	\bibitem[{\citenamefont{Lu and Kofke}(1999)}]{Lu1999}
	\bibinfo{author}{\bibfnamefont{N.}~\bibnamefont{Lu}} \bibnamefont{and}
	\bibinfo{author}{\bibfnamefont{D.~A.} \bibnamefont{Kofke}},
	\bibinfo{journal}{J. Chem. Phys} \textbf{\bibinfo{volume}{111}},
	\bibinfo{pages}{4414} (\bibinfo{year}{1999}).
	
	\bibitem[{\citenamefont{Radmer and Kollman}(1997)}]{Radmer1997}
	\bibinfo{author}{\bibfnamefont{R.~J.} \bibnamefont{Radmer}} \bibnamefont{and}
	\bibinfo{author}{\bibfnamefont{P.~A.} \bibnamefont{Kollman}},
	\bibinfo{journal}{J. Comput. Chem} \textbf{\bibinfo{volume}{18}},
	\bibinfo{pages}{902} (\bibinfo{year}{1997}).
	
	\bibitem[{\citenamefont{Wu and Kofke}(2005)}]{Wu2005}
	\bibinfo{author}{\bibfnamefont{D.}~\bibnamefont{Wu}} \bibnamefont{and}
	\bibinfo{author}{\bibfnamefont{D.~A.} \bibnamefont{Kofke}},
	\bibinfo{journal}{J. Chem. Phys} \textbf{\bibinfo{volume}{123}},
	\bibinfo{pages}{084109} (\bibinfo{year}{2005}).
	
	\bibitem[{\citenamefont{Sugita et~al.}(2000)\citenamefont{Sugita, Kitao, and
			Okamoto}}]{Sugita2000}
	\bibinfo{author}{\bibfnamefont{Y.}~\bibnamefont{Sugita}},
	\bibinfo{author}{\bibfnamefont{A.}~\bibnamefont{Kitao}}, \bibnamefont{and}
	\bibinfo{author}{\bibfnamefont{Y.}~\bibnamefont{Okamoto}},
	\bibinfo{journal}{J. Chem. Phys} \textbf{\bibinfo{volume}{113}},
	\bibinfo{pages}{6042} (\bibinfo{year}{2000}).
	
	\bibitem[{\citenamefont{Earl and Deem}(2005)}]{Earl2005}
	\bibinfo{author}{\bibfnamefont{D.~J.} \bibnamefont{Earl}} \bibnamefont{and}
	\bibinfo{author}{\bibfnamefont{M.~W.} \bibnamefont{Deem}},
	\bibinfo{journal}{Phys. Chem. Chem. Phys} \textbf{\bibinfo{volume}{7}},
	\bibinfo{pages}{3910} (\bibinfo{year}{2005}).
	
	\bibitem[{\citenamefont{Kofke}(2002)}]{Kofke2002}
	\bibinfo{author}{\bibfnamefont{D.~A.} \bibnamefont{Kofke}},
	\bibinfo{journal}{J. Chem. Phys} \textbf{\bibinfo{volume}{117}},
	\bibinfo{pages}{6911} (\bibinfo{year}{2002}).
	
	\bibitem[{\citenamefont{Predescu et~al.}(2004)\citenamefont{Predescu, Predescu,
			and Ciobanu}}]{Predescu2004}
	\bibinfo{author}{\bibfnamefont{C.}~\bibnamefont{Predescu}},
	\bibinfo{author}{\bibfnamefont{M.}~\bibnamefont{Predescu}}, \bibnamefont{and}
	\bibinfo{author}{\bibfnamefont{C.~V.} \bibnamefont{Ciobanu}},
	\bibinfo{journal}{J. Chem. Phys} \textbf{\bibinfo{volume}{120}},
	\bibinfo{pages}{4119} (\bibinfo{year}{2004}).
	
	\bibitem[{\citenamefont{Kone and Kofke}(2005)}]{Kone2005}
	\bibinfo{author}{\bibfnamefont{A.}~\bibnamefont{Kone}} \bibnamefont{and}
	\bibinfo{author}{\bibfnamefont{D.~A.} \bibnamefont{Kofke}},
	\bibinfo{journal}{J. Chem. Phys} \textbf{\bibinfo{volume}{122}},
	\bibinfo{pages}{206101} (\bibinfo{year}{2005}).
	
	\bibitem[{\citenamefont{Rathore et~al.}(2005)\citenamefont{Rathore, Chopra, and
			de~Pablo}}]{Rathore2005}
	\bibinfo{author}{\bibfnamefont{N.}~\bibnamefont{Rathore}},
	\bibinfo{author}{\bibfnamefont{M.}~\bibnamefont{Chopra}}, \bibnamefont{and}
	\bibinfo{author}{\bibfnamefont{J.~J.} \bibnamefont{de~Pablo}},
	\bibinfo{journal}{J. Chem. Phys} \textbf{\bibinfo{volume}{122}},
	\bibinfo{pages}{024111} (\bibinfo{year}{2005}).
	
	\bibitem[{\citenamefont{Predescu et~al.}(2005)\citenamefont{Predescu, Predescu,
			and Ciobanu}}]{Predescu2005}
	\bibinfo{author}{\bibfnamefont{C.}~\bibnamefont{Predescu}},
	\bibinfo{author}{\bibfnamefont{M.}~\bibnamefont{Predescu}}, \bibnamefont{and}
	\bibinfo{author}{\bibfnamefont{C.~V.} \bibnamefont{Ciobanu}},
	\bibinfo{journal}{J. Phys. Chem. B} \textbf{\bibinfo{volume}{109}},
	\bibinfo{pages}{4189} (\bibinfo{year}{2005}).
	
	\bibitem[{\citenamefont{Nadler and Hansmann}(2007)}]{Nadler2007}
	\bibinfo{author}{\bibfnamefont{W.}~\bibnamefont{Nadler}} \bibnamefont{and}
	\bibinfo{author}{\bibfnamefont{U.~H.} \bibnamefont{Hansmann}},
	\bibinfo{journal}{Phys. Rev. E} \textbf{\bibinfo{volume}{76}},
	\bibinfo{pages}{065701} (\bibinfo{year}{2007}).
	
	\bibitem[{\citenamefont{Chodera and Shirts}(2011)}]{Chodera2011Replica}
	\bibinfo{author}{\bibfnamefont{J.~D.} \bibnamefont{Chodera}} \bibnamefont{and}
	\bibinfo{author}{\bibfnamefont{M.~R.} \bibnamefont{Shirts}},
	\bibinfo{journal}{J. Chem. Phys.} \textbf{\bibinfo{volume}{135}},
	\bibinfo{pages}{194110} (\bibinfo{year}{2011}).
	
	\bibitem[{\citenamefont{Dirks et~al.}(2012)\citenamefont{Dirks, Xu, and
			Shaw}}]{Dirks2012}
	\bibinfo{author}{\bibfnamefont{R.~M.} \bibnamefont{Dirks}},
	\bibinfo{author}{\bibfnamefont{H.}~\bibnamefont{Xu}}, \bibnamefont{and}
	\bibinfo{author}{\bibfnamefont{D.~E.} \bibnamefont{Shaw}},
	\bibinfo{journal}{J. Chem. Theory Comput.} \textbf{\bibinfo{volume}{8}},
	\bibinfo{pages}{162} (\bibinfo{year}{2012}).
	
	\bibitem[{\citenamefont{MacCallum et~al.}(2018)\citenamefont{MacCallum,
			Muniyat, and Gaalswyk}}]{Maccallum2018}
	\bibinfo{author}{\bibfnamefont{J.~L.} \bibnamefont{MacCallum}},
	\bibinfo{author}{\bibfnamefont{M.~I.} \bibnamefont{Muniyat}},
	\bibnamefont{and} \bibinfo{author}{\bibfnamefont{K.}~\bibnamefont{Gaalswyk}},
	\bibinfo{journal}{J. Phys. Chem. B} \textbf{\bibinfo{volume}{122}},
	\bibinfo{pages}{5448} (\bibinfo{year}{2018}).
	
	\bibitem[{\citenamefont{Large and Sivak}(2019)}]{Large2019}
	\bibinfo{author}{\bibfnamefont{S.~J.} \bibnamefont{Large}} \bibnamefont{and}
	\bibinfo{author}{\bibfnamefont{D.~A.} \bibnamefont{Sivak}},
	\bibinfo{journal}{J. Stat. Mech.: Theory Exp.}
	\textbf{\bibinfo{volume}{2019}}, \bibinfo{pages}{083212}
	(\bibinfo{year}{2019}).
	
	\bibitem[{\citenamefont{Imparato et~al.}(2007)\citenamefont{Imparato, Peliti,
			Pesce, Rusciano, and Sasso}}]{Imparato2007}
	\bibinfo{author}{\bibfnamefont{A.}~\bibnamefont{Imparato}},
	\bibinfo{author}{\bibfnamefont{L.}~\bibnamefont{Peliti}},
	\bibinfo{author}{\bibfnamefont{G.}~\bibnamefont{Pesce}},
	\bibinfo{author}{\bibfnamefont{G.}~\bibnamefont{Rusciano}}, \bibnamefont{and}
	\bibinfo{author}{\bibfnamefont{A.}~\bibnamefont{Sasso}},
	\bibinfo{journal}{Phys. Rev. E} \textbf{\bibinfo{volume}{76}},
	\bibinfo{pages}{050101} (\bibinfo{year}{2007}).
	
	\bibitem[{\citenamefont{Sellentin et~al.}(2014)\citenamefont{Sellentin,
			Quartin, and Amendola}}]{Sellentin2014}
	\bibinfo{author}{\bibfnamefont{E.}~\bibnamefont{Sellentin}},
	\bibinfo{author}{\bibfnamefont{M.}~\bibnamefont{Quartin}}, \bibnamefont{and}
	\bibinfo{author}{\bibfnamefont{L.}~\bibnamefont{Amendola}},
	\bibinfo{journal}{Mon. Not. R. Astron. Soc.} \textbf{\bibinfo{volume}{441}},
	\bibinfo{pages}{1831} (\bibinfo{year}{2014}).
	
	\bibitem[{\citenamefont{Goldberg and Bacon}(2005)}]{Goldberg2005}
	\bibinfo{author}{\bibfnamefont{D.~M.} \bibnamefont{Goldberg}} \bibnamefont{and}
	\bibinfo{author}{\bibfnamefont{D.~J.} \bibnamefont{Bacon}},
	\bibinfo{journal}{Astrophys. J} \textbf{\bibinfo{volume}{619}},
	\bibinfo{pages}{741} (\bibinfo{year}{2005}).
	
	\bibitem[{\citenamefont{Bacon et~al.}(2006)\citenamefont{Bacon, Goldberg, Rowe,
			and Taylor}}]{Bacon2006}
	\bibinfo{author}{\bibfnamefont{D.~J.} \bibnamefont{Bacon}},
	\bibinfo{author}{\bibfnamefont{D.~M.} \bibnamefont{Goldberg}},
	\bibinfo{author}{\bibfnamefont{B.~T.~P.} \bibnamefont{Rowe}},
	\bibnamefont{and} \bibinfo{author}{\bibfnamefont{A.~N.}
		\bibnamefont{Taylor}}, \bibinfo{journal}{Mon. Not. R. Astron. Soc.}
	\textbf{\bibinfo{volume}{365}}, \bibinfo{pages}{414} (\bibinfo{year}{2006}).
	
\end{thebibliography}
\end{document}